\documentclass[a4paper,fleqn,usenatbib]{mnras}

\usepackage{graphicx}	
\usepackage{amsmath}	
\usepackage{amssymb}	
\usepackage{epstopdf}
\usepackage{subeqnarray}
\usepackage[para]{threeparttable} 
\usepackage{rotating}
\usepackage{multirow}
\usepackage[usenames, dvipsnames]{color}

\def\vsini{V_\mathrm{eq} \sin i} 
\def\kms{\mathrm{km.s}^{-1}}
\def\phidiff{\phi_\mathrm{diff}}
\def\Rsun{\mathrm{R}_{\sun}}

\def\Msun{\mathrm{M}_{\sun}}

\def\Tmean{\overline{T}_\mathrm{eff}}
\def\diameq{\diameter_\mathrm{eq}}

\title[DI of the rapid rotator Regulus ]{Differential interferometry of the rapid rotator Regulus \thanks{Based on observations performed at the European Southern Observatory, Chile under ESO AMBER Visitor mode program IDs 092.D-0342}}

\author[M. Hadjara et al.]{
M. Hadjara,$^{1,2,3}$\thanks{E-mail: Massinissa.Hadjara@oca.eu}
R. G. Petrov,$^{2}$
S. Jankov,$^{4}$
P. Cruzal\`ebes,$^{2}$
A. Spang,$^{2}$
and S. Lagarde$^{2}$
\\
$^{1}$Instituto de Astronom\`ia, Universidad Cat\`olica del Norte, Av. Angamos 0610 Antofagasta, Chile\\
$^{2}$Universit\'e de Nice–Sophia Antipolis (UNS), Centre National de la Recherche Scientifique (CNRS), Observatoire de la C\^ote d’Azur (OCA),\\
 Laboratoire J. L. Lagrange, UMR 7293,Campus Valrose, 06108 Nice Cedex 2, France\\ 
$^{3}$Centre de Recherche en Astronomie, Astrophysique et G\'{e}ophysique (CRAAG), Route de l'Observatoire, B.P. 63, Bouzareah, 16340,\\ Alger, Algeria\\
$^{4}$Astronomical Observatory, Volgina 7, PO Box 74, 11060 Belgrade, Serbia
}

\date{Accepted 2018 July 12. Received 2018 July 1; in original form 2017 October 9}

\pubyear{2018}

\begin{document}
\label{firstpage}
\pagerange{\pageref{firstpage}--\pageref{lastpage}}
\maketitle

\begin{abstract}
We analyse interferometric data obtained for Regulus with AMBER (Astronomical Multi- BEam combineR) at high spectral resolution ($\lambda/\delta\lambda \approx 12000$) across the Br$\gamma$ spectral line. The study of the photocentre displacement allows us to constrain a large number of stellar parameters -- equatorial radius $R_{\rm eq}$, equatorial velocity $V_{\rm eq}$, inclination $i$, rotation-axis position angle $PA_{\rm rot}$, and flattening -- with an estimation of gravity-darkening coefficient $\beta$ using previously published theoretical results. We use the Simulation Code of Interferometric-observations for ROtators and CirCumstellar Objects (SCIROCCO), a semi-analytical algorithm dedicated to fast rotators. We chose Regulus because it is a very well-known edge-on star, for which an alternative approach is needed to check the previously published results. Our analysis showed that a significant degeneracy of solution is present.\\
By confronting the results obtained by differential interferometry with those obtained by conventional long-base interferometry, we obtain similar results (within the uncertainties), thereby validating our approach, where $V_{eq}$ and $i$ are found separately. From the photocentre displacement, we can independently deduce $PA_{rot}$. We use two minimization methods to restrict observed stellar parameters via a fast rotator model: a non-stochastic method ($\chi^2$ fit) and a stochastic one (Markov Chain Monte Carlo method), in order to check whether the correct global minimum is achieved particularly with respect to the degeneracies of the gravity darkening parameter $\beta$, where we demonstrate, using a quantitative analysis of parameters, that the estimate of $\beta$ is easier for stars with an inclination angle of around $45^\circ$.\\
\end{abstract}

\begin{keywords}
methods: numerical -- methods: observational -- techniques: high angular resolution -- techniques: interferometric -- stars: individual: Regulus.
\end{keywords}
\section{Introduction }
\label{introduction}
\subsection{Optical interferometry of rapid rotators}
\label{Opt-Interf-rapid_rot}

Stellar rotation was measured for the first time by interferometry, from the photocentre displacements by \cite{sl94}, on the slow rotator Aldebaran, which was observed in 1988 at OHP (Observatoire de Haute-Provence) through the 152cm telescope by the Speckle Differential Interferometry method. Results obtained by interferometry on the fast rotators were summarized by \cite{2011SerAJ.183....1J} \& \cite{2012A&ARv..20...51V}.
The extreme stellar flattening induced by the rotation was measured by interferometry by \cite{2003A&A...407L..47D} on Achernar ($R_{eq}/R_{pol}=1.56\pm0.05$), using VLTI/VINCI (Very Large Telescope Interferometer/VLT INterferometer Commissioning Instrument) with a dense ($u,v$) coverage.
The first image reconstruction of the surface of a fast rotator, showing the gravity darkening effect \citep{1924MNRAS..84..665V, 1924MNRAS..84..684V} was on Altair \citep{2007Sci...317..342M} from CHARA (Center for High Angular Resolution Astronomy) observations, inferring several fundamental parameters: inclination; position angle; effective temperature; and polar and equatorial radii.

Inspired by the early works of \cite{1975ApJ...196L..71L}, \cite{1982AcOpt..29..361B} proposed the Differentiel Speckle Interferometry technique, using the chromatic displacement of the speckle photocentre given by the first-order term of the phase of the spatial Fourier transform of the sky brightness according to the MacLaurin series \citep{2001A&A...377..721J}. This technique has been extended to a wider range of wavelengths and applied to long-baseline interferometry by \cite{1988ESOC...29..235P, 1989dli..conf..249P} who established the fundamentals of the Differential Interferometry (DI) technique. This allowed, for the first time, to measure simultaneously the angular separation and the radial velocity difference of the two stellar components of the binary Capella \citep{1992ASPC...32..477P} using the photocentre as a function of the wavelength.

The combination of high spatial and high spectral resolution allows us to measure physical properties of fast rotators beyond the diffraction limit, as shown by \cite{2012A&A...545A.130D} and \cite{2014A&A...569A..45H} who used the differential phases from AMBER/VLTI (Astronomical Multi-BEam combineR). Indeed, AMBER \citep{2007A&A...464....1P} is a spectro-interferometric instrument specifically designed to go well beyond the resolution limit \citetext{e.g., \citealt{2007A&A...464...59M,2009A&A...498L..41L}}.

Optical interferometry provides several types of measures, as the absolute visibility, differential visibility \& closure phase \citep{2007A&A...464....1P}, but in this paper we focus only on differential phase and vectorial photocentre displacement. 

\subsection{The fast rotator Regulus }
\label{Reg}

$\alpha$ Leo A (HR 3982, HD 87901) one of the brightest stars of the sky, is a binary system which brighter primary component is referred as Regulus throughout this paper. Regulus is an edge-on and flattened nearby star which is in rapid rotation. In the following we summarize the spectrophotometric and the interferometric information of our target separately.

\subsubsection{Information from spectroscopy and photometry}
\label{spec-phot_info}

With magnitude $V=1.40$ \citep{2009ApJ...694.1085V}, Regulus has been identified as a fast rotator by \cite{1954ApJ...119..146S}, who determined by spectroscopy its high rotationnal velocity $\vsini=352\pm7.5\kms$ i.e. $96\%$ of its critical velocity.

$\alpha$ Leo is a multiple stellar system composed of at least two binaries. The A component of the system ($\alpha$ Leo A, HD 87901) has been recently discovered to be a spectroscopic binary. The brighter companion (Regulus) was classified as a main sequence B7V star by \cite{1953ApJ...117..313J}, and more recently as a sub-geant B8IV star by \cite{2003AJ....126.2048G} of mass $\sim$ 4 $\Msun$ \cite[][and references therein]{2011ApJ...732...68C}.
\cite{2008ApJ...682L.117G} argue that the fainter companion of $\alpha$ Leo A is probably a white dwarf or a M4 V star of mass $\sim$ 0.3 $\Msun$ and an orbital period of 40.11 days, and that the magnitude difference of the fainter component with respect to Regulus in the K band is close to $\Delta m_K\approx10$ ($\sim 10^{-4}$ of flux ratio) and  6 ($\sim 4\times 10^{-3}$ of flux ratio), for the cases of a white dwarf and an M4 V star companion, respectively. Thus, the flux of the companion has no influence on our analysis (Br$_\gamma$ and adjacent continuum), as well as on the interferometry presented by \cite{2005ApJ...628..439M} and \cite{2011ApJ...732...68C}.
For this reason only an extraordinary activity of the fainter star, reflected as a several order of magnitude enhancement of the Br$_\gamma$ emission, could eventually affect our results.
$\alpha$ Leo A has a companion which is in fact a system of two other components (B and C) which together form a binary system \citep{2005ApJ...628..439M}.
The B component ($\alpha$ Leo B; HD 87884) is an $\sim$ 0.8 $\Msun$ star of spectral type K2V while the C component is a very faint M4V star with a mass of $\sim$ 0.2 $\Msun$. 
The Washington Double Star Catalog \citep{2001AJ....122.3466M} lists a D component, also having a common proper motion with the system and a separation of $\approx$3.6$\arcmin$ from the A component while the B-C subsystem is located $\approx$ 3$\arcmin$ from the A component.

\cite{2008Ap&SS.318...51I} studied the possible correction of the Keplerian period due to the quadrupole mass moment induced by the oblateness of Regulus. Although this correction could be measured in principle, the total uncertainty in the Keplerian period \citep[0.02 days from][]{2008ApJ...682L.117G} due to the errors in the systems parameters (mostly in the velocity semiamplitude and in the mass of Regulus) is larger than the correction by about two orders of magnitude.
Its distance is $d=24.3\pm0.2pc$ according to \cite{2007ASSL..350.....V} and $d=23.759\pm0.045pc$ according to \cite{2009ApJ...694.1085V}. Its mass is $M=4.15\pm0.06\Msun$ from the $Y^2$ stellar evolution model \citep{2001ApJS..136..417Y, 2003ApJS..144..259Y, 2004ApJS..155..667D}, $3.66^{+0.79}_{-0.28}\Msun$ from \cite{2011ApJ...732...68C} according to the oblateness mass method of \cite{2009ApJ...701..209Z} and $3.80\pm0.6\Msun$ from \cite{1990A&AS...85.1015M} according to the evolutionary tracks of \cite{1989A&A...210..155M}. Its age is estimated between 150 Myr \citep{2001A&A...379..162G} and 1 Gyr \citep{2009ApJ...698..666R}. Its effective temperature $T_{\rm eff}$ is $12460\pm200$K according to \cite{1990A&AS...85.1015M} and $11960\pm80$K according to \cite{2003AJ....126.2048G}. Its metallicity $[M/H]$ is $0.0$ according to \cite{2003AJ....126.2048G}.

\subsubsection{Interferometric observations of Regulus}
\label{Interfero-info}
The first interferometric observations of this star were done with the Narrabri Intensity Interferometer by \cite{1974MNRAS.167..121H}. Because of the poor ($u,v$)-plane coverage, only information about its size could be obtained, with an equatorial angular diameter $\diameq=1.32\pm0.06 mas$. Using the CHARA array observations in the K-band, \cite{2005ApJ...628..439M} measured for the first time the inclination of its rotation axis $i=90^\circ$ $^{+0}_{-15}$, and characterized other physical parameters such as: rotation-axis position angle $PA_{rot}=265.5\pm2.8^\circ$; rotational equatorial velocity $V_{eq}=317^{+3}_{-85}\kms$; fractional rotational velocity $\frac{V_{eq}}{V_{crit}}=0.86\pm0.03$; equatorial and polar radii $R_{eq}=4.16\pm0.08\Rsun$ \& $R_{pol}=3.15\pm0.06\Rsun$; equatorial and polar effective temperatures $T_{eq}=10314\pm1000K$ \& $T_{pol}=15400\pm1400K$; mass $M$; luminosity $L$; gravity darkening coefficient $\beta$ \citep[defined as $T_{\rm eff} \propto g_{\rm eff}^{\beta}$ by][where $T_{\rm eff}$ and $g_{\rm eff}$ are local effective temperature and gravity respectively]{1924MNRAS..84..665V}; distance $d$ and interstellar extinction $A_v$. More recently, \cite{2011ApJ...732...68C} used the CHARA/MIRC (Michigan Infra-Red Combiner) instrument to produce maps of $\alpha$ Leo, in the H-band, and deduced inclination angle $i=86.3^\circ$$^{+1.0^\circ}_{-1.6^\circ}$ and gravity darkening coefficient $\beta=0.188^{+0.012}_{-0.029}$ which is consistent (within the uncertainties) with the results of \cite{2005ApJ...628..439M}.
So, interferometry revealed that Regulus is an edge-on star with an inclination angle of $i\sim90^\circ$ and rotationnally flattened with an oblateness ratio (equatorial-to-polar radii minus 1; $R_{eq}/R_{pol}-1$) reported between $0.325\pm0.036$ (angular diameter $\diameq=1.65\pm0.02mas$) and $0.307\pm0.030$ ($\diameq=1.61_{-0.02}^{0.03}mas$) \citep{2005ApJ...628..439M, 2011ApJ...732...68C}.

Table.~\ref{table2} summarizes the fundamental parameters of Regulus.

\begin{table}
\begin{minipage}{87mm}
\caption{Fundamental stellar parameters of $\alpha$ Leo found in the literature.} \label{table2}
\centering
\begin{threeparttable}
\centering
\begin{tabular}{l|c}
\hline \hline
Parameter & Value\\
\hline \hline
Angular diameter ($\diameq$) & $1.65\pm0.02mas$ (1)\\
                             & $1.61_{-0.02}^{0.03}mas$ (2)\\
\hline
Oblateness ratio ($R_{eq}/R_{pol}-1$) & $0.325\pm0.036$ (1)\\
                                    & $0.307\pm0.030$ (2)\\
\hline
Distance (d) & $24.3\pm0.2pc$ (3)\\
             & $23.759\pm0.045pc$ (4)\\
\hline
         & $4.15\pm0.06\Msun$ (5)\\
Mass (M) & $3.66^{+0.79}_{-0.28}\Msun$ (2)\\
         & $3.80\pm0.6\Msun$ (6)\\
\hline
Age & 50 - 200 Myr (7)\\
    & $\geq$ 1 Gyr (8)\\
\hline
Eff. temperature ($T_{eff}$) & $12460\pm200$K (6)\\
                             & $11960\pm80$K (9)\\
\hline
Metallicity ($[M/H]$) & $0.0$ (9)\\
\hline
Rotation-axis position angle $PA_{rot}$ & $265.5\pm2.8 ^\circ$ (10) \\
                                        & $258^{+2}_{-1}$$^\circ$ (11) \\
\hline
Rotation-axis inclination angle $i$ & $90^{+0}_{-15}$$^\circ$ (10) \\
                                        & $85.3^{+1}_{-1.6}$$^\circ$ (11) \\
\hline \hline

\end{tabular}
\begin{tablenotes}
		\footnotesize
		$(1)$ \citet{2005ApJ...628..439M}
		$(2)$ \citet{2011ApJ...732...68C}\\
		$(3)$ \citet{2007ASSL..350.....V}
		$(4)$ \citet{2009ApJ...694.1085V}\\
		$(5)$ \citet{2004ApJS..155..667D}
		$(6)$ \citet{1990A&AS...85.1015M}\\
		$(7)$ \citet{2001A&A...379..162G}
		$(8)$ \citet{2009ApJ...698..666R}\\
		$(9)$ \citet{2003AJ....126.2048G}
		$(10)$ \citet{2005ApJ...628..439M}\\
		$(11)$ \citet{2011ApJ...732...68C}
\end{tablenotes}
\end{threeparttable}
\end{minipage}
\end{table}

\subsection{Structure of the article}
\label{Struct_paper}

In this paper, we describe the differential interferometry with high spectral resolution observations ($R\simeq12000$) in the K band of the rapid rotator Regulus and we focuss on the parameters that can extracted from the photocentre displacement, eventually in combination with the broadened spectral line profile. We compare this in detail with the parameters obtained from a broad band interferometric images and we discuss the values that are inferred or improved by the comparison and then the combination of both techniques.

The present paper is organized as follows:
\begin{itemize}
\item In Sec.~\ref{obsdatared}, we present the observations and the data reduction of Regulus.
\item In Sec.~\ref{geom_params}, we study the photocentre displacement of our target, where we deduce $PA_{\rm rot}$ directly from the observed photocentres displacements.
\item In Sec.~\ref{Modeling}, we present the model that was used in order to interpret our measurements and discuss the constraints that they give on the gravity darkening parameter $\beta$ of Regulus.
\item In Sec.~\ref{Fitting}, we fit the fundamental parameters of Regulus, using a non-stochastic method ($\chi^2$); and a stochastic one (Markov Chain Monte Carlo method).
\item In Sec.~\ref{discus}, we summarize the computed accuracy limits that we could achieve with the quality of our data and we discuss the probability spaces of the couple $(\beta,i)$ of Regulus.
\item In Sec.~\ref{conclusions}, we analyse the results and open the discussion to a broader study of fast-rotating stars observed with VLTI-AMBER by DI.
\end{itemize}

\section{Observations and data reduction }
\label{obsdatared}

Regulus was observed with the AMBER/VLTI instrument located at Cerro Paranal, Chile, with the Auxiliary Telescopes. The fringes were stabilized using FINITO (Fringe-tracking Instrument of NIce and TOrino) \citep{2012SPIE.8445E..1KM} as a fringe tracker, allowing us to use a Detector Integration Time DIT=3sec for 20 exposures. The observations have been performed using the high spectral resolution mode of AMBER ($\lambda/\delta\lambda\approx12000$). Table~\ref{table1} provides the observation log of Regulus.

The differential phase $\phidiff$ obtained from the data reduction algorithm is related to the object's Fourier phase $\phi_{\rm obj}$ by \citetext{e.g., \citealt{2006EAS....22..379M}, \citeyear{2011A&A...526A.107M}}:
\begin{equation}
\phidiff(u,v)=\phi_{\rm obj}(u,v)-a(u,v)-b(u,v)/\lambda,
\label{eq8}
\end{equation}
where the spatial frequency coordinates $u$ and $v$ depend on the wavelength $\lambda$, the projected baseline length $B_{\rm proj}$ and the baseline position angle $PA$ (from north to east; $u = B_{\rm proj}\sin(PA)/\lambda$ and $v = B_{\rm proj}\cos(PA)/\lambda$). The parameters $a$ and $b$ correspond to an offset and a slope, given in appropriate units.

\begin{figure}
\centering
 \includegraphics[width=1.\hsize,draft=false]{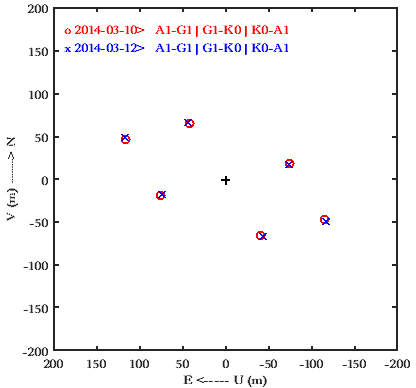}
 \caption{($u,v$) coverage for our VLTI/AMBER observations of Regulus.} \label{UV_Coverage_Reg}
\end{figure}

The corresponding $(u,v)$ coverage is shown in Fig.\ref{UV_Coverage_Reg}, where the $(u,v)$ plane is spanned over $\sim$1.5 h/night. Note the rather poor sampling of the Fourier space. According to the Table~\ref{table1}, the $(u,v)$ points for the date 2014-03-10 are represented by red circles and for the date 2014-03-12 by blue crosses.

\begin{table}
\centering
\caption{VLTI/AMBER observations of Regulus and its calibration stars using AT triplet A1-G1-K0. Note that the detector integration time (DIT)=3 s, frame number per exposure NDIT=20 and number of all used exposures NEXP=32 for the first night (2014-03-10) and 22 for the second night (2014-03-12).} \label{table1}
\begin{tabular}{cccc}
\hline \hline
Object & Date \& time & Baseline length & Baseline PA\\
 & & $\textit{B}_{\rm proj}$(m) & \textit{PA}$(^\circ)$\\

 \hline \hline

60 Cnc & 2014-03-10 T03:09 & 75,81,128 & 103,34,67 \\
Regulus & 2014-03-10 T03:48 & 78,77,125 & 104,32,68 \\
w Cen & 2014-03-10 T04:30 & 73,87,129 & 88,14,47 \\
w Cen & 2014-03-10 T04:44 & 74,87,129 & 90,16,50 \\

\hline
$\epsilon$ Cma & 2014-03-12 T02:15 & 75,87,116 & 124,36,76 \\
Regulus & 2014-03-12 T03:59 & 76,79,127 & 104,33,68 \\
w Cen & 2014-03-12 T04:45 & 75,87,129 & 92,17,51 \\
$\iota$ Cen & 2014-03-12 T07:17 & 80,88,126 & 113,30,69 \\

\hline \hline

\end{tabular}
\end{table}

Data have been reduced using version 3.0.9 of the \textit{amdlib} software \citep{2009A&A...502..705C, 2007A&A...464...29T}. We adopt a mild frame selection based on fringe signal to noise (S/N) and geometric flux to noise thresholds greater than unity. Our dataset includes stellar spectrum; differential visibilities; differential phases; and closure phases.


\subsection{Spectrum }
\label{Res_Spec}

The high spectral resolution mode of AMBER leads to a velocity resolution of $\simeq25\kms$. The projected equatorial rotational velocities $\vsini$ above $\sim150 \kms$ of Regulus ensures that the $Br\gamma$ line, is sampled by $6$ spectral channels. Thus rotation effects should be taken into account when modeling phase signatures.

\begin{figure}
\centering
 \includegraphics[width=1.\hsize,draft=false]{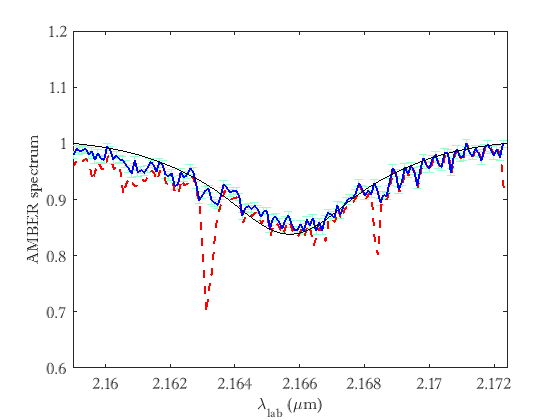}
 \caption{AMBER spectrum of Regulus in the Br$_\gamma$ line. The dashed thick red curve is the raw spectrum showing two telluric lines. The full thick blue curve is the Regulus spectrum, with its error bars in green. The thin dark line represents our best model, which is discussed in section \ref{chi2}.} \label{Spec_Reg_RP_WEB}
\end{figure}

Figure \ref{Spec_Reg_RP_WEB} shows the normalized observed flux of Regulus as observed (dashed red curve) and after correction (in blue line). The smooth black curves superimposed on the observations is our best model that is discussed in section \ref{chi2}. The normalized observed flux was corrected by removing the both strong tellurics and doing a wavelength calibration. We converted the wavelength scale to the laboratory frame using Reguluss heliocentric velocity of $5.90\pm2.40\kms$ measured by \citet{2012AstL...38..331A} and the \textit{IRAF} package \footnote{IRAF is distributed by the National Optical Astronomy Observatories, which are operated by the Association of Universities for Research in Astronomy, Inc., under cooperative agreement with the National Science Foundation.}, in order to convert the observed velocity to the heliocentric frame ($RV=-10.195\kms$, the average of the both nights ; $-9.69\kms$  for 10/03/2014 \& $-10.70\kms$ for 12/03/2014.

\subsection{Visibilities }
\label{Res_Vis}

\begin{figure*}
\centering
 \includegraphics[width=0.4\hsize,draft=false]{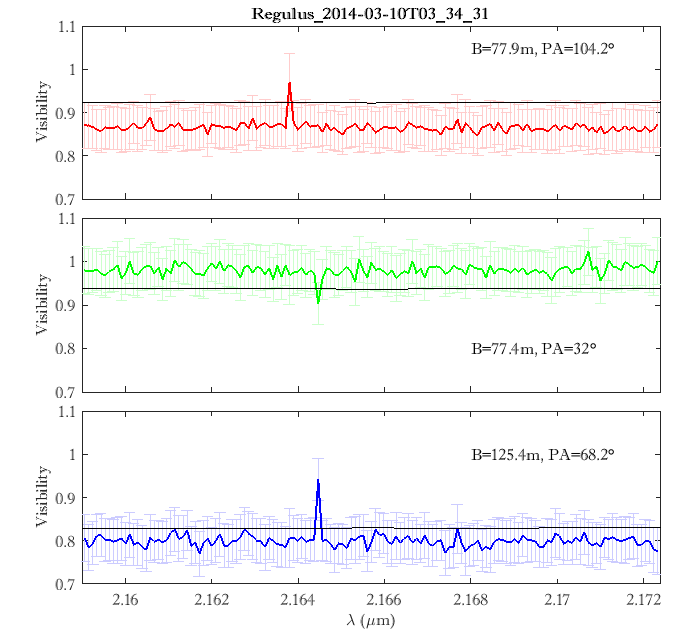}
 \includegraphics[width=0.4\hsize,draft=false]{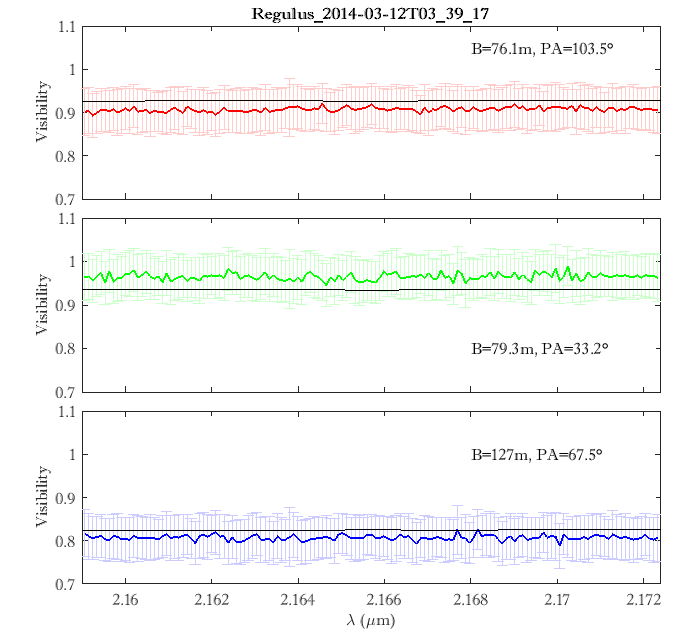}
 \caption{The six observed visibilities of Regulus as coloured thick lines, with uncertainties overplotted from the modelling visibilities as black thin lines. The equatorial radius of our best model corresponds to $R_{\rm eq}=4.16\pm0.24\Rsun$, the polar radius is $R_{\rm pol}=3.08\pm0.27\Rsun$ and $PA_{\rm rot}=251\pm2^\circ$ (see Sec.~\ref{chi2}).} \label{Reg_vis}
\end{figure*}

Figure \ref{Reg_vis} shows the observed visibilities which are superimposed by our best model (black line) that is discussed in section \ref{chi2}. The second baseline ($B_{proj}\approx 78m$ \& $PA\approx 32^\circ$), which are the closest to the polar direction of Regulus, are those with the visibility closest to 1.

The baselines values show that the angular resolution $\lambda/B$ is always larger than 3.5 mas. As the largest diameter of Regulus is smaller than 1.7 mas, the source is not enough resolved for image reconstruction, as confirmed by our closure phases which are equal to zero within the noise, as shown in Fig.~\ref{Reg_cphi} below.

\begin{figure}
\centering
 \includegraphics[width=1.\hsize,draft=false]{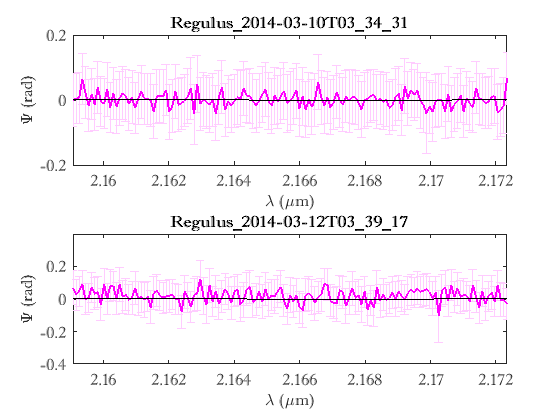}
 \caption{Both observed closure phases of Regulus as coloured thick lines with uncertainties overplotted from the modelling as black thin lines and with the same parameters as Fig.~\ref{Reg_vis} (see Sec.~\ref{chi2}).} \label{Reg_cphi}
\end{figure}

The longest projected baseline on the polar direction is $\sim74$m. With our 0.04 error on the visibility, an angular diameter of 1.1 mas is measured with typically a 0.7 mas accuracy that means that it can be anything smaller than 1.8 mas (or even 2.5 mas at 2 $\sigma$ level). On the equatorial direction, the longest projected baseline is $\sim125$m and a 1.6 mas diameter is measured with an accuracy of typically 0.16 mas.  We see that the absolute visibilities on our VLTI baselines in the K band can be used to estimate the largest diameter of Regulus with an accuracy comparable to previously published results but is useless to constrain the oblateness with any useful accuracy. In other words, our absolute visibility measurements cannot resolve the oblateness of Regulus, or its position angle and cannot give any access to the consequences of its rapid rotation. This is confirmed by our measurements with V=1 within noise on at least one baseline in each observation.

Regulus is so marginally resolved and its visibilities are constant and flat (the average visibility of the six ($u,v$) points is around 0.9) within the noise. The visibilities are flat, because we are in high spectral resolution (R$\sim$12000); the variation of the wavelength range is so weak that the slopes of the visibilities are indistinguishable within the noise. Qualitatively, our wavelength range is between 2.15 and 2.17 $\mu m$. The radius of Regulus is $\approx$ 1.63 mas and our largest baseline is of 125m. With a simple uniform disk model we obtain a visibility of 0.830 for $\lambda = 2.15\mu m$ and visibility = 0.833 for $\lambda = 2.17\mu m$ (so, a difference of 0.3\% in terms of visibility between the first and the last wavelength). 

\subsection{Differential phase, vectorial photocentre displacement and closure phase }
\label{Res_Phi}

The differential phase is the variation of the phase through a spectral line with respect to the phase in the continuum forced to zero (see Eq.~\ref{eq8}). On resolved sources, the differential phase boosts the imaging capability \cite{2011A&A...526A.107M}. On non-resolved sources, the differential phase is proportional to the photocentre variation of the source with wavelength $\epsilon(\lambda)$, with respect to the photocentre of the source in the continuum $\epsilon(\lambda_{\rm c})$, as follow:

\begin{equation}
\phi_{ij}(\lambda)=2\pi\stackrel{\rightarrow}{\epsilon}(\lambda)\stackrel{\rightarrow}{B_{ij}}/\lambda,
\label{eq1}
\end{equation}

As soon as it is measured on two baselines ${\vec B}_{ij}$ ($i \ne j$), it yields the vectorial photocentre displacement $\stackrel{\rightarrow}{\epsilon}(\lambda)-\stackrel{\rightarrow}{\epsilon}(\lambda_c)$. As the absolute photocentre of the source is unknown, we decide by convention that the photocentre of the source in the continuum (or in the reference channel) is the origin of the coordinate system and that $\stackrel{\rightarrow}{\epsilon}(\lambda_c)=0$. To simplify the equations, in the following we shall just use $\stackrel{\rightarrow}{\epsilon}(\lambda)$ but we have to remember that it is defined in a coordinate system with the photocentre in the spectral reference channel at its origin.
On non-resolved sources the differential phase decreases like $\frac{\diameq}{\lambda/B}$. Like spectro-astrometry on single apertures \citep{2015A&A...579A..48W} it yields the photocentre variation vector on source much smaller than the diffraction limit. This makes the differential phase optical interferometry measure with the highest "super-resolution" potential. For example, with a differential phase accuracy of the order of 5 milliradian (mrad) we have a photocentre displacement accuracy of 4 micro-arcesecond ($\mu$as) with AMBER on the VLTI \citep[achieved by][on Fomalhaut]{2009A&A...498L..41L}.
The vectorial photocentre displacement yields the position angle of the rotation axis \citep[e.g. Fomalhaut by][]{2009A&A...498L..41L}, angular sizes and rotation velocities. It allows to separate spatially and spectrally the different components of the source with different spectral characteristics or radial velocities. This has been achieved first on non-resolved slow rotators by \cite{sl94} and on fast rotators by \cite{2003A&A...407L..47D}.

The visibilities are quasi flat also on the Br$_{\gamma}$ line and obviously all closure phases are equal to zero.
In this case and according to \cite{1989dli..conf..249P} (and after that by \cite{2003A&A...400..795L} using the moment of the flux distribution which is $\approx$ object size/spatial resolution), who have demonstrated that the phase ($\phi $) is proportional to the first order of this quantity, the visibility modulus ($1-|V|^2$) is proportional to the second order and the closure phase ($\Psi$) to the $3^{rd}$ one.

Thus, only the differential phase can give useful angular resolution information. So all the baselines can be projected on two photocentre coordinates, as soon as we have baselines in at least two different directions. It means that, we obtain $\epsilon_{\alpha}$ in the right ascension (East to West) direction and $\epsilon_{\delta}$ in the declination (South to North) direction, from an average of the projections of all baselines on these two directions. The contribution of each baseline is weighted by its signal-to-noise ratio (SNR).

We are therefore in the situation when all differential phases are given by the equation \ref{eq1} can be projected on two orthogonal axes or plotted as a vector (Fig.~\ref{Regulus_PArot}).

\section{Independent determination of $PA_{rot}$}
\label{geom_params}

\begin{figure*}
\centering
\includegraphics[width=0.48\hsize,draft=false]{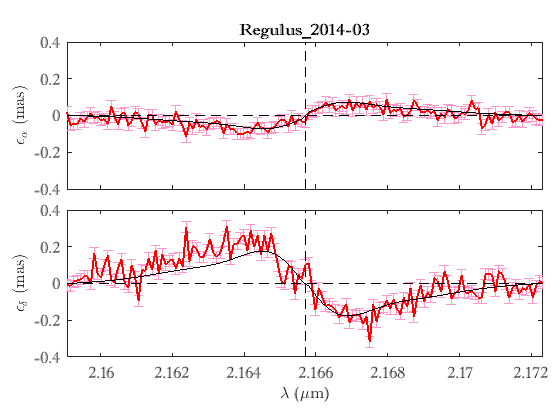}
\includegraphics[width=0.48\hsize,draft=false]{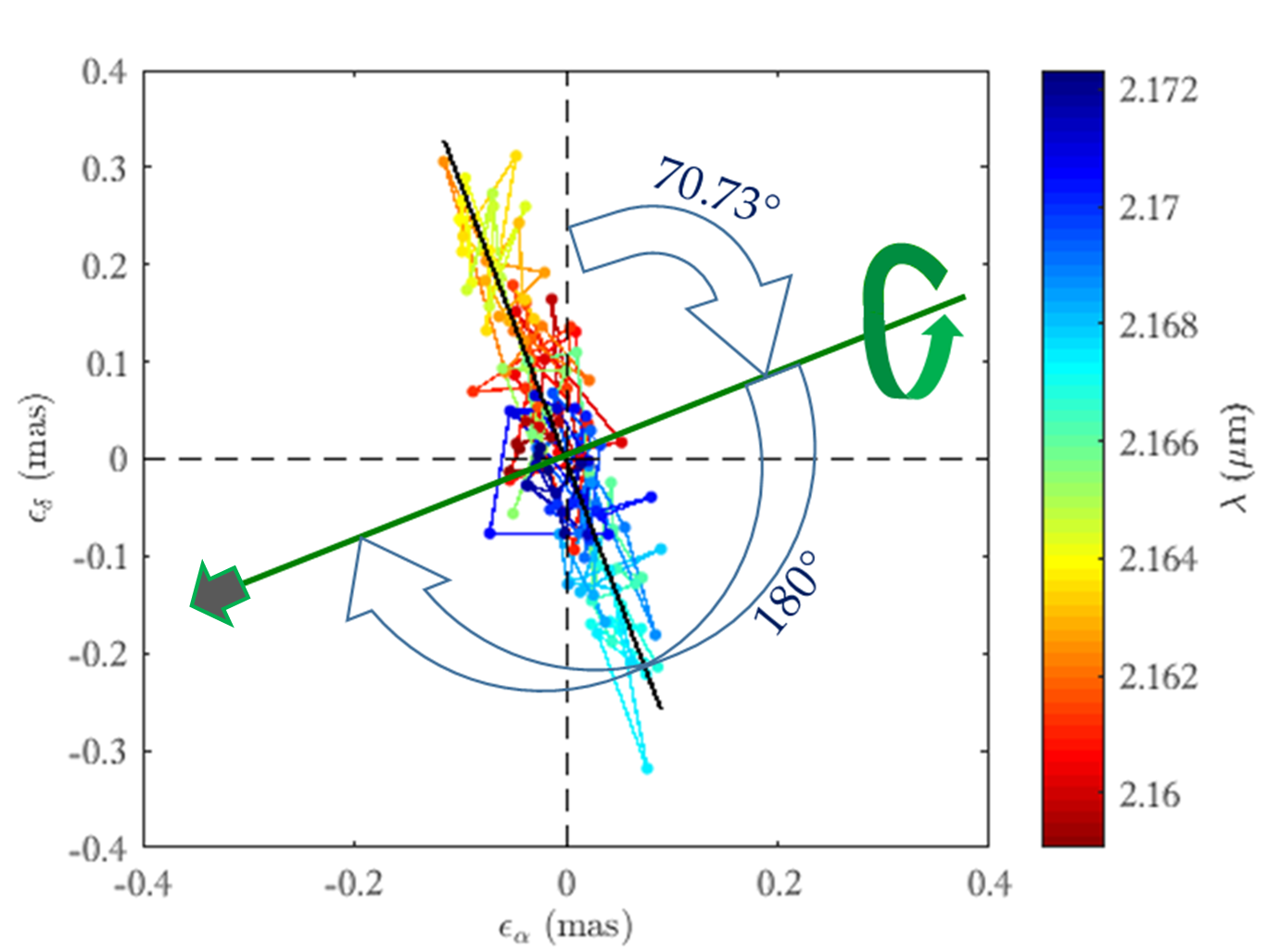}
\caption{\textbf{Left:} the perpendicular right ascension and declination photocentre displacements $(\epsilon_{\alpha},\epsilon_{\delta})$ as red thick curves for the observed data with uncertainties. The smooth thin black curves superimposed on the observations show the best-fitting $\phidiff$ as discussed in Sec.~\ref{chi2}. The RMS error per spectral channel has been measured in the continuum outside the spectral line and found to be $\sim 30$ $\mu as$ on any projection. The two perpendicular dashed lines represent the zero-point for the photocentre displacement axis and the central wavelength ($\lambda=2165.7 nm$) of the $Br_{\gamma}$ line. \textbf{Right:} the vectorial photocentre displacement on the sky. Each point represents a wavelength, as indicated by the colour bar. The black line is the fit through all points that indicates the direction of the equator and the green perpendicular line represents the rotation axis, which can be deduced directly from the angle $PA_{rot}$. The grey arrow with the green outline represents the apparent stellar pole. If we strictly apply the definition cited above $PA_{rot}=250.73^\circ\pm3^\circ$.}
\label{Regulus_PArot}
\end{figure*}

In this section, we deduce independently the $PA_{rot}$ of the star directly and only from observed photocentre displacements, as it has been done in the past ; first by \cite{1992ASPC...32..477P} on the binary Cappella, by \cite{sl94} on the slow rotator Aldebaran, then by \cite{2009A&A...498L..41L} on Fomalhaut and its circumstellar debris disk.
$PA_{\rm rot}$ is defined as the angle from north to east until the stellar rotation axis at the visible stellar pole. With this definition and by a simple linear fit of our data ($\epsilon_{\alpha}=a_1\epsilon_{\delta}+a_2$), we find $PA_{rot}$ as the slope of the line; $PA_{rot}=\pi+\arctan(a_1)=250.73^\circ\pm3^\circ$ ($+\pi$ because of the first visible stellar pole which is to west of the SED). 
This value is close to the previous results of $PA_{rot}$ \protect\citep{2005ApJ...628..439M, 2011ApJ...732...68C}. Fig.~\ref{Regulus_PArot} shows the photocentre displacement of Regulus, which is in the same equatorial direction of our target.

The vectorial representation in Fig.~\ref{Regulus_PArot} directly gives the position angle of the rotation vector. 
Differential Interferometry allows to find the exact orientation of the rotation vector. A comparison with the motion of sources close to Regulus (its companions in the first place) could give constraints on Regulus history but this is out of the scope of this paper.

\begin{figure}
\centering
\includegraphics[width=1.\hsize,draft=false]{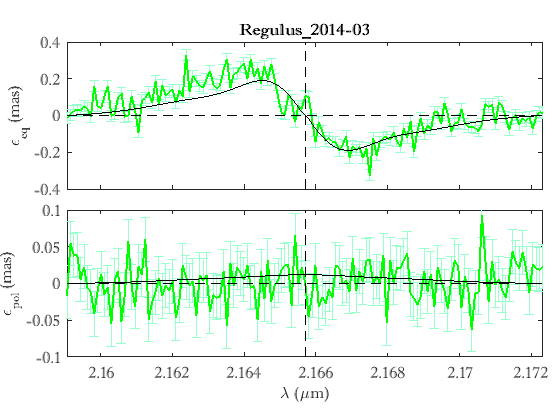}
\caption{The perpendicular equatorial-polar photocentre displacements $(\epsilon_{\rm eq},\epsilon_{\rm pol})$ shown as green thick curves for the observed data with uncertainties, both corresponding to the six VLTI/AMBER achieved ($u,v$) coverage points of Regulus, around Br $\gamma$ at two different observing times, for each time. The smooth thin black curves superimposed on the observations are the best-fitting $\phidiff$ as discussed in Sec.~\ref{chi2}. The two perpendicular dashed lines represent the zero-point  for the photocentre displacement axis and the central wavelength ($\lambda=2165.7 nm$) of the $Br_{\gamma}$ line.}
\label{Chi2_Fit_Regulus_Npix=256_Config1}
\end{figure}

Once $PA_{\rm rot}$ is known, using the elementary coordinate frame rotation rules we can deduce the equatorial and polar photocentre displacements $(\epsilon_{eq},\epsilon_{pol})$ from the photocentre displacements $(\epsilon_{\alpha},\epsilon_{\delta})$. Figure \ref{Chi2_Fit_Regulus_Npix=256_Config1} shows the observed photocentre displacements $(\epsilon_{eq},\epsilon_{pol})$ with our best model (black line) that is discussed in section \ref{chi2}.
The used uncertainties for the photocentre displacements are the root mean square (RMS) of those of $\phidiff$ at the continuum. We used the RMS because the Data Reduction Software of \textit{amdlib} calculates the uncertainties from the differential piston, which is centered on central wavelength ($\lambda_c$) and are higher on the continuum than on the central line.

\section{Modelling rapid rotators and their photocentre displacement}
\label{Modeling}

In addition to the $PA_{rot}$ angle of the rotation vector estimate we need a model of Regulus to interpret our measurements.

\subsection{SCIROCCO}
\label{SCIROCCO}

\begin{figure*}
\centering
 \includegraphics[width=0.32\hsize,draft=false]{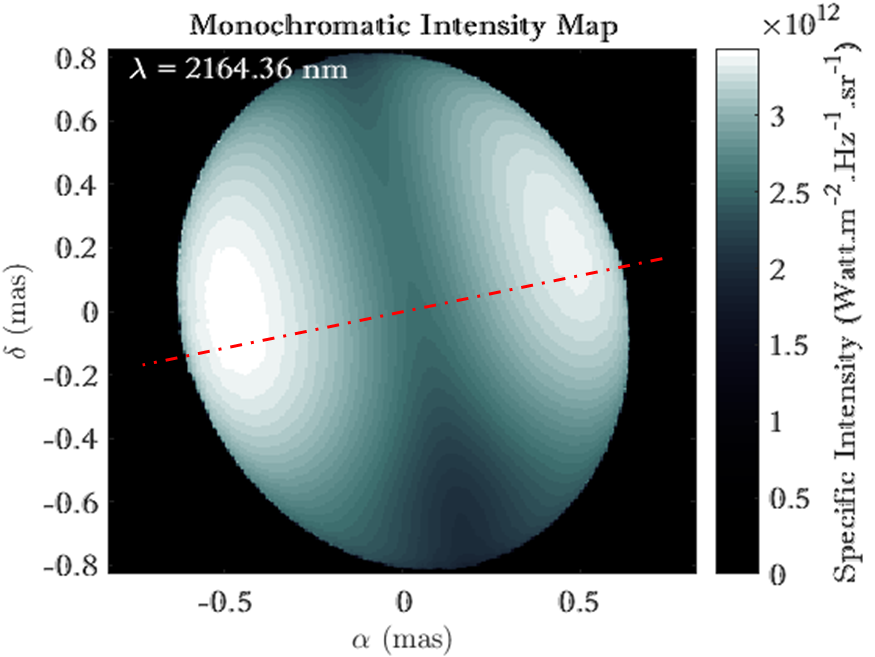}
 \includegraphics[width=0.32\hsize,draft=false]{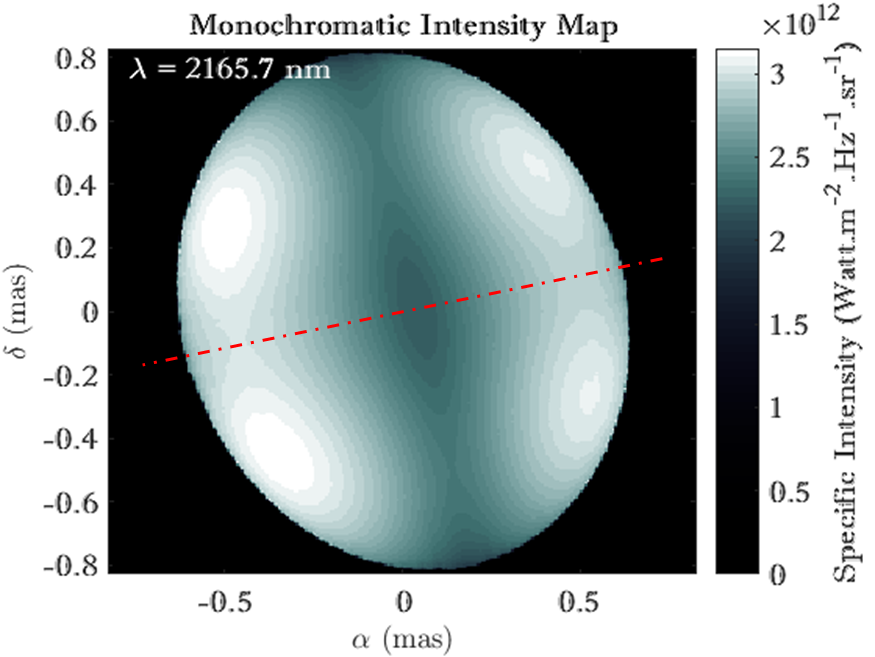}
 \includegraphics[width=0.32\hsize,draft=false]{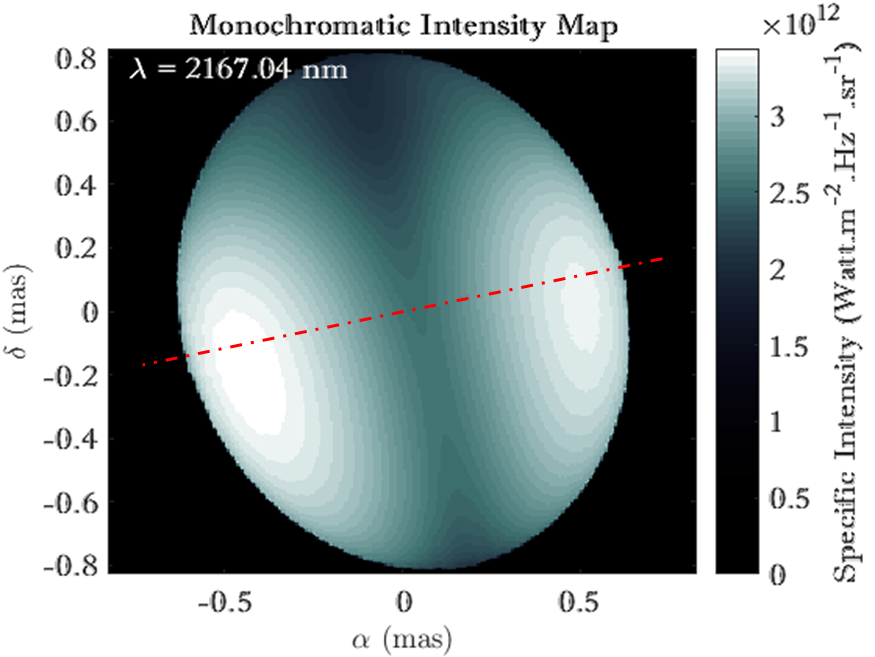}
 \caption{Three monochromatic intensity maps of Regulus from our simulator SCIROCCO, with physical parameters as discussed in Sec.~\ref{chi2}, which are close to those of \citet{2011ApJ...732...68C}. The three wavelengths are in the Br~$\gamma$ line and represent radial velocities of $-185.6$, $0$ and $+185.6$ $\kms$, from left to right. The red dashed line represents the rotation axis of the star.} \label{Reg_Scirocco}
\end{figure*}

To interpret the $\phidiff$ observations, we use the semi-analytical model for fast rotators: SCIROCCO {S}imulation {C}ode of {I}nterferometric-observations for {RO}tators and {C}ir{C}umstellar {O}bjects. This code written in \texttt{Matlab}\footnote{MATrix LABoratory}, allows to compute monochromatic intensity maps of uniformly rotating, flattened, and gravity-darkened stars using semi-analytical approach. SCIROCCO, which is a parametric description of the velocity field, the intensity map and line profile modelisation, in each point (latitude, longitude), allow us to obtain directly from the modeled specific intensity maps, on the photospheric lines, the spectro-interferometric observables as spectra and photocentres, and by using Fourier transformations, the visibility amplitudes, phases, and closure phases. SCIROCCO is described in detail in \cite{2014A&A...569A..45H, Massi2015}.
The Fig. \ref{Reg_Scirocco} shows the modeled monochromatic intensity maps, by our model for the fast rotators study, for a given Doppler shift at three wavelengths around the Br~$\gamma$ line, adopting the stellar parameters given by our best-fit, as discussed in Sec~\ref{chi2}, and which are close to those of \cite{2011ApJ...732...68C} i.e. $R_{\rm eq}=4.16 \Rsun$;  $V_{\rm eq}=350\kms$; $i=86.4^\circ$; and $PA_{\rm rot}=251^\circ$. For this figure we use the gravity darkening coefficient estimated following the theoretical method of \citet{2011A&A...533A..43E} approach (see Sec.~\ref{chi2}), where $\beta\approx0.17$.

\subsection{Fixed parameters}
\label{Fix_param}

\begin{figure}
\centering
 \includegraphics[width=1.\hsize,draft=false]{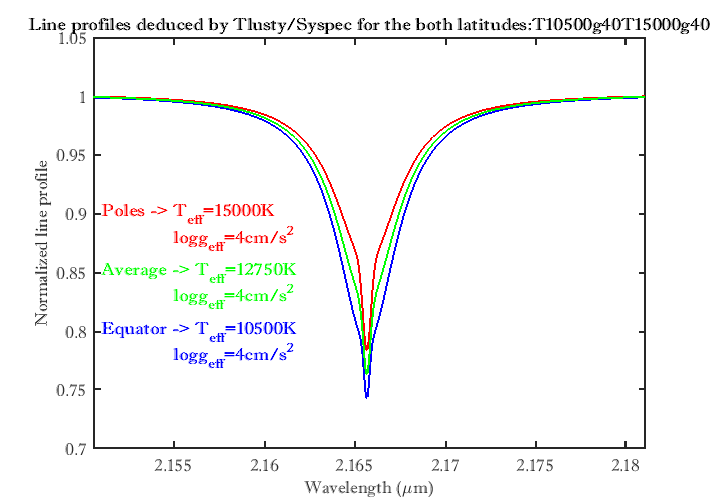}
 \caption{Latitude dependence of our Br$_\gamma$  line profiles.} \label{Reg_lp}
\end{figure}

The fixed parameters that we use for our modeling are:

\begin{itemize}
  \item{Local velocity field.}
  \begin{itemize}
    \item{Microturbulence: any case lower than $10 \kms$ has no impact as the resulting line broadening is much smaller than our spectral PSF (Point Spread Function) width of $25$ $\kms$.
    We chose the Regulus's microturbulent velocity value as the solar one (i.e. $2$ $\kms$) by default, in order to simulate the limb-darkening effect. This choice has no effect on our final results, because Regulus is marginally resolved and in addition Regulus's line profile is too large ($\overline{FWHM}=26\Delta\lambda$) and shallow ($|\overline{Amp}|=0.24$) (see Fig.~\ref{Reg_lp}) to be sensitive to an accurate value of the limb-darkening coefficient in general and to the microturbulent velocity in particular.}
    \item{Differential rotation. We can introduce it, but here we have neglected it because it introduces signatures that disappear in the noise.}
    \end{itemize}
  \item{Line profile. Unlike \cite{1995A&AS..109..401C}, we use a different line strength for each latitude ($\theta$) of the star fixed by the couples of latitudinal temperature and latitudinal surface gravity $[T_{\theta},\log g_{\theta}]$ from Kurucz/Synspec (synthetic spectrum) stellar atmosphere modelling.}
\end{itemize}

Figure \ref{Reg_lp} shows the local Br$_\gamma$ line profile representation for a star, from Kurucz/Synspec model at 3 different latitudes; with $[T_{pol},\log g_{\rm pol}]=[15000K,4cm/s^2]$ at the poles (red line), $[T_{\rm eq},\log g_{\rm eq}]=[10500K,4cm/s^2]$ at the equator  (blue) and the average ($[12750K,4cm/s^2]$, in green). The polar line profiles have less amplitude than the equatorial one.

The relevant fixed parameters of our model for rapid-rotating star Regulus are:

\begin{itemize}
  \item Distance $d=23.759\pm0.446pc$ given by \cite{2009ApJ...694.1085V} from HIPPARCOS data \citep{1997A&A...323L..49P}.
  \item Mass $M=3.8 \pm 0.57 \Msun$ given by \cite{1990A&AS...85.1015M}.
  \item Surface mean effective temperature $\Tmean=12500 K$ given by \cite{1990A&AS...85.1015M}. Using the spectral energy distributions (SEDs) from our model, with this $\Tmean$, we obtain the apparent magnitudes $m_V=1.4\pm0.1$ and $m_K=1.6\pm0.1$ which are consistent, within the uncertainties, with those found in the Strasbourg astronomical Data Center\footnote{Available at \url{http://cdsweb.u-strasbg.fr/}}: $m_V=1.40\pm0.05$ and $m_K=1.62\pm0.05$ \citep{2009ApJ...694.1085V}.
  \item The limb darkening is fixed, assuming the Claret function \citep{2000A&A...363.1081C}, by the following parameters:
  \begin{itemize}
    \item micro-turbulent velocity $VT=2\kms$ (solar standard $VT$),
    \item surface mean effective surface gravity $\log g =4 cm/s^2$,
    \item metallicity $[Fe/H]^e=0.0$ given by \cite{2011ApJ...732...68C} and
    \item spectral filter in the K band.
  \end{itemize}
\end{itemize}

For our model, we have selected the fixed parameters which we use in this paper. In the litterature one can find slightly different values for this fixed parameters. We have checked that these changes have no impact on our interferometric measurable, at our level of accuracy, nor to the parameters that we extract from a fit of our measures, as it is discussed in Sec.~\ref{chi2}.

Concerning the gravity darkening coefficient, it is indirectly estimated $\beta$, following \citet{2011A&A...533A..43E}, in function of the polar and equatorial radius ratio, where; 
\begin{equation}
\beta=\frac{1}{4}-\frac{1}{3}(1-\frac{R_{\rm pol}}{R_{\rm eq}}).
\label{eqbeta}
\end{equation}
The formula that links $R_{\rm eq}$, $R_{\rm pol}$ and $V_{\rm eq}$ is given by \cite{2014A&A...569A..45H, Massi2015} and \cite{2002A&A...393..345D} and shown in Appendix \ref{scirocco_ded}.

\subsection{Sensitivity of the photocentre displacement to $\beta$}
\label{sensitiv_eps}

Before describing the global fits of $R_{\rm eq}$, $V_{\rm eq}$, $i$, $PA_{\rm rot}$ and, tentatively $\beta$ we will use our model to illustrate the sensitivity of our differential photocentre measures to $\beta$. 
\begin{figure*}
\centering
\includegraphics[width=0.48\hsize,draft=false]{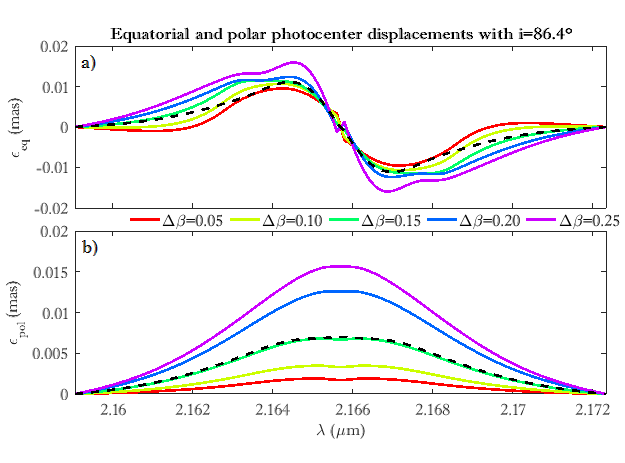}
\includegraphics[width=0.48\hsize,draft=false]{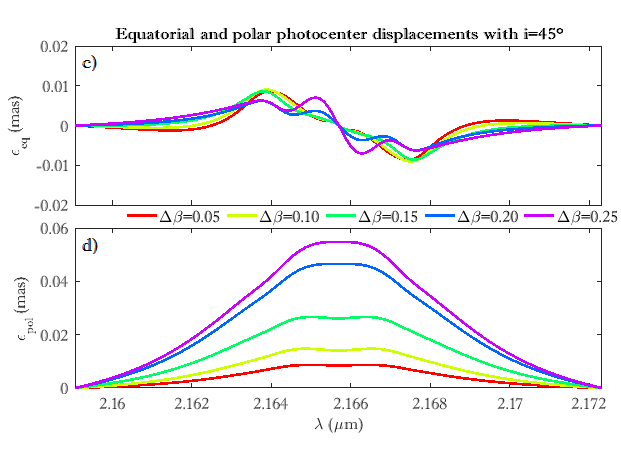}
\caption{(a) and (b) Effect of changes in $\beta$ on the equatorial and polar photocentre displacements, compared with the effect of changes in radius or inclination. In colour we plot $\epsilon(\lambda,\beta)-\epsilon(\lambda,\beta=0)$ for the parameters of "param$\_$set3". The dashed dark line in the $\epsilon_{\rm eq}$ figure (a) shows the effect of a 6\% radius variation that has the same total amplitude as a $\beta = 0.15$ variation. In the  $\epsilon_{\rm pol}$ figure (b), the same line shows the effect of a $8.4^\circ$ variation of the inclination i that has the same total amplitude as a $\beta = 0.15$ variation. 
(c) and (d) Same as (a) and (b), but with an inclination $i=45^\circ$.}
\label{horseshoes}
\end{figure*}

Figures \ref{horseshoes}a and \ref{horseshoes}b shows respectively the sensitivity of the equatorial and polar photocentre displacement to a change in the parameter $\beta$. With the parameters of our best model (see Sec.~\ref{chi2}), which are close of those used by \cite{2011ApJ...732...68C}, we plot the differences between $\epsilon(\lambda)$ for $\beta$ and $\epsilon(\lambda)$ for $\beta=0$; i.e. $\Delta\stackrel{\rightarrow}{\epsilon}(\lambda,\beta)=\stackrel{\rightarrow}{\epsilon}(\lambda,\beta)-\stackrel{\rightarrow}{\epsilon}(\lambda,\beta=0)$.

We expected a maximum signature of $\beta$ in the rotation axis direction signal $\Delta\epsilon_{\rm pol}(\lambda)$ because the gravity darkening introduces a dissymmetry between the northern and southern hemispheres, if the star is not exactly equator-on. Indeed, we see a relatively regular change of the average amplitude of $<\Delta\epsilon_{\rm pol}(\lambda)>_\lambda$ of the order of 8 $\mu as$ for $\Delta\beta$=0.25. 
As $<\Delta\epsilon_{\rm pol}(\lambda)>_\lambda$ can be estimated with an accuracy of about 2.5 $\mu as$, this would yield an accuracy $\sigma_\beta\approx\frac{0.25}{8}\times2.5=0.08$ if $\beta$ is the unique unknown. However, we see that the polar photocentre displacement $\Delta\epsilon_{\rm pol}(\lambda)$ will not allow to separate the effects of $\beta$ and $i$: for example, a variation of $\beta$ from 0 to 0.15 gives almost exactly the same effect on $\Delta\epsilon(\lambda)$ than a change in inclination $i=8.4^\circ$, as shown with the dashed line of figure \ref{horseshoes}b ($\epsilon_{\rm pol}(\lambda,\beta=0.15, i = 78^\circ, R_{\rm eq}= 4.16\Rsun) - \epsilon_{\rm pol}(\lambda,\beta=0.15, i = 86.4^\circ, R_{\rm eq}= 4.16\Rsun)$).
Surprisingly, $\beta$ has also a strong impact on the equatorial photocentre displacement $\Delta\epsilon_{\rm eq}(\lambda)$ that can be explained by the fact that the extra polar brightness enhances the weight of parts of the star with small and intermediate radial velocities with respect to the high radial velocity regions. 
Changes in the stellar radius have a direct influence on $\epsilon_{\rm eq}(\lambda)\propto R_{\rm eq}$. The signature of a change in radius for a given $\beta$ and of changes in $\beta$ for a given radius are different. $R_{\rm eq}$ changes only the global amplitude of $\epsilon_{\rm eq}(\lambda)$ while $\beta$ changes the shape of $\epsilon_{\rm eq}(\lambda)$ with, for example, a strong change in the wings of $\epsilon_{\rm eq}(\lambda)$ when $\beta$ varies from 0 to 0.25. 
We have therefore to consider two cases: If $\epsilon_{\rm eq}(\lambda)$ is known, our measure of $<\left|\epsilon_{\rm eq}(\lambda)\right|>_\lambda$ would yield $\sigma_\beta\approx 0.08$. 
If we consider that $R_{\rm eq}$ is unknown and must be estimated from our data, then the specific signature of a variation in $\beta$ is reduced. An uncertainty of $\frac{\sigma_R}{R}\leq 1.5\%$ on $R_{\rm eq}$ as from \cite{2005ApJ...628..439M} \& \cite{2011ApJ...732...68C}, would have almost no impact and allow our data to constrain $\sigma_\beta\leq 0.1\%$ that would be far from decisive. If we have the uncertainty $\frac{\sigma_R}{R}\leq 6\%$ that results from our differential interferometry data alone, as we will see it in the next section, the specific signature of a change in $\beta$, i.e. for example the difference between the variation on $\epsilon_{\rm eq}(\lambda)$ when $\beta$ varies from 0.05 to 0.25 and this due to a 6\% change in radius, averaged over $\lambda$ is reduced to less than 3 $\mu$as, yielding an uncertainty $\sigma_\beta\geq 0.2$ that is useless to constrain the modeling of the Von Zeipel effect. The dashed line of Fig.~\ref{horseshoes}a shows the effect of angular diameter changes of 6\% on $\epsilon_{\rm eq}(\lambda)$ ($\epsilon_{\rm eq}(\lambda,\beta=0.15, i = 86.4^\circ, R_{\rm eq}= 4.41\Rsun) - \epsilon_{\rm eq}(\lambda,\beta=0.15, i = 86.4^\circ, R_{\rm eq}= 4.16\Rsun)$) which has the same extrema values than the curve for a change of $\Delta\beta=0.15$.

Figs.~\ref{horseshoes}c-d, like Figs.~\ref{horseshoes}a-b but with $i=45^\circ$, shows a less $\epsilon_{\rm eq}$ in amplitude and width because of the $\vsini$.
At this inclination, where the gravity darkening effect is more pronounced over $\epsilon_{\rm pol}$, it will be easier for us to determine $\beta$ of Regulus in the same conditions. Indeed, between $i=90^\circ$ and $i=45^\circ$, $\epsilon_{\rm pol}$ amplitude wins a factor of $\sim$ 3 \& the impact of $\beta$ is more distinguishable, where $\epsilon_{\rm pol}$ is larger than its $\sigma_\epsilon=30\mu as$, when $\epsilon_{\rm pol}$ is drowned in its noise for $i=90^\circ$ (Fig.~\ref{Chi2_Fit_Regulus_Npix=256_Config1}). What it means that the $\beta$ estimation method using the photocentre displacements that we propose here is supposed to work better with fast rotators with inclination around of $45^\circ$, despite their angular resolution and/or their SNR.

The physical characteristics $[T_{eff},\log g]$, from equator to pole, of the line profiles corresponding for each used $\beta$ in this paper, are summarized in the Table \ref{tab_prob_beta}.

\begin{table}
\centering
\caption{Physical characteristics $[T_{eff},\log g]$, from equator to pole, of the line profile corresponding to each $\beta$.} \label{tab_prob_beta}
\begin{tabular}{cccc}
\hline \hline
 & Range  & Real & Kurucz/Tlusty\\
$\beta$ & of & $[T_{eff},\log g]$ & $[T_{eff},\log g]$\\
 & $\beta$ & $(K,cm/s^{2})$ & $(K,cm/s^{2})$ \\

\hline \hline
0.00 & - & [12500, 4.0] & [12500, 4.0]\\
\hline
 & & [11946, 3.78] & [12000, 4.0]\\
0.05 & 0.05-0.07 & to & to \\
 & & [13054, 4.04] & [13000, 4.0]\\
\hline
 & & [11367, 3.78] & [11500, 4.0]\\
0.10 & 0.08-0.12 & to & to \\
 & & [13633, 4.04] & [14000, 4.0]\\
\hline
 & & [10762, 3.78] & [10750, 4.0]\\
0.15 & 0.13-0.17 & to & to \\
 & & [14238, 4.04] & [14000, 4.0]\\
\hline
 & & [10130, 3.78] & [10000, 4.0]\\
0.20 & 0.18-0.22 & to & to \\
 & & [14870, 4.04] & [15000, 4.0]\\
\hline
 & & [9471, 3.78] & [9500, 4.0]\\
0.25 & 0.23-0.25 & to & to \\
 & & [15529, 4.04] & [15000, 4.0]\\
\hline \hline

\end{tabular}
\end{table}

Note also that $\epsilon_{pol}$ is not zero at $\beta =0$, because of the $\vsini$ effect. Indeed, $V_{\rm eq}$ \& $i$ which remains the same (i.e. $350\kms$ \& 86.4$^\circ$), produce an asymmetrical SED in the polar direction, which makes a none zero $\epsilon_{pol}$ at $\beta =$ 0 even with a 1D fixed line profile ($[\Tmean, \log g_{\rm eff}] = [12500K, 4cm/s^{-2}]$).

\section{Fitting the fundamental parameters of Regulus}
\label{Fitting}

\subsection{$\chi2$ minimization}
\label{chi2}

In order to deduce the best parameters fitting differential phase data, we perform a $\chi^2$ minimization\footnote{The discription of this method was explained well in \cite{2014A&A...569A..45H}} for the $\phidiff$ data. We use here the corresponding photocentre displacements $(\epsilon_{\rm eq},\epsilon_{\rm pol})$. The used uncertainties in this minimization are those of the $\phidiff$.

We use four model-fitting parameters sets ("param$\_$set"). The free parameters are: $R_{\rm eq}$ (equatorial radius); $V_{\rm eq}$ (equatorial velocity); $i$ (inclination angle); and $PA_{\rm rot}$ (rotation-axis position angle).
Because $PA_{\rm rot}$ is determined well by the method of the photo-center displacement slope that we shown above in Sec.~\ref{geom_params}, we start with a model-fitting to deduce three free parameters; $R_{\rm eq}$, $V_{\rm eq}$ \& $i$ (with $PA_{rot}$ fixed). We also checked the real independence of $PA_{\rm rot}$ to the best solution using a model-fitting with four free parameters; $R_{\rm eq}$, $V_{\rm eq}$, $i$ \& $PA_{\rm rot}$.

In our study, where we change the value of $\beta$ (fixed/indirectly estimated by the Eq.~\ref{eqbeta}), The couples equatorial-polar effective temperature and effective surface gravity $[T_{eq},\log g_{\rm eq}]$ \& $[T_{\rm pol},\log g_{\rm pol}]$ needed for constructing our three dimensional (3D) line profile are deduced by running one time the $\chi^2$ fitting, of our model on the observation data, with a fixed line profile $[T_{\rm eff},\log g_{\rm eff}]$.

Parameters set 1 ("param$\_$set1") represents the most complete SCIROCCO modeling of Regulus for 3 free parameters $R_{\rm eq}$, $V_{\rm eq}$ \& $i$ (with $PA_{\rm rot}$ fixed), and gravity darkening coefficent $\beta$ theoretically estimated from the Eq.~\ref{eqbeta}, an analytic 3D Kurucz/Synspec line profile, and latitudinal limb darkening (depending on $\theta$). The parameters set 2 ("param$\_$set2") is similar to the "param$\_$set1" but with $\beta$ fixed to the value of $0.25$. 

The parameters set 3 ("param$\_$set3") represents the most complete SCIROCCO modeling of Regulus for 4 free parameters $R_{\rm eq}$, $V_{\rm eq}$, $i$ \& $PA_{\rm rot}$. The parameters set 4 ("param$\_$set4") is similar to the "param$\_$set3" but with $\beta$ fixed to $0.25$.

We show here only the results of the most complete model ("param$\_$set3"), because we found that all the four parameter sets results are similar within the uncertainties. 
All the best-fit values of all the parameter sets are summarized in Table \ref{tab_param_Reg_plus} in the appendix \ref{all_fit_res}.
The similarity of the $PA_{\rm rot}$ results between "param$\_$set3\&4" and parameter sets 1 \& 2 confirm the independence of $PA_{\rm rot}$. And the similarity of the results between all the parameter sets (with respect to the uncertainties), where $\beta$ is fixed and indirectly deduced, confirm the difficulty to constrain $\beta$ with our current data.

Table~\ref{tab_param_Reg1} summarize the results of the "param$\_$set3" which has been choosed as the reference "param$\_$set" model-fitting with the MCMC uncertainties (whose the justification of this choice is  well explained in the Sec.~\ref{MCMC}), where $R_{\rm eq}=4.16\pm0.24\Rsun$, $V_{\rm eq}=350\pm22\kms$, $i=86.4\pm6.3^\circ$ \& $PA_{\rm rot}=251\pm2^\circ$.
The comparison of our results with those found in the literature (in Tab.~\ref{tab_param_Reg1}) confirms that the use of differential phases alone for the $\chi^2$ minimization is largely sufficient to constrain the fundamental stellar parameters, as has been done in the past for Achernar by \cite{2012A&A...545A.130D} and for Achernar, Altair, $\delta$ Aquilae \& Fomalhaut by \cite{2014A&A...569A..45H}.

Figures \ref{Regulus_PArot} (left) \& \ref{Chi2_Fit_Regulus_Npix=256_Config1} (above -Sec.~\ref{geom_params}-) shows the best-fit on the both photocentre displacements $(\epsilon_{eq},\epsilon_{pol})$  \& $(\epsilon_{\alpha},\epsilon_{\delta})$ obtained with "param$\_$set3". In the Fig.~\ref{Chi2_Fit_Regulus_Npix=256_Config1} we observe that the equatorial photocentre displacement $\epsilon_{eq}$ is more important than the polar one $\epsilon_{pol}$. Indeed, the rotation is in the equatorial direction and the asymmetry in the modeling $\epsilon_{pol}$, which is quasi flat, corresponds to the combination of the inclination angle $i=86.4^\circ$ and the Von Zeipel effect (those last produce a small asymetry, around the $Br_{\gamma}$ line, on $\epsilon_{\rm pol}$. this asymetry is obvious in the modeling $\epsilon_{\rm pol}$ and buried in the noise of the observations). Where $\epsilon_{\rm eq}$ start from the first wavelength at zero for increasing before go to zero at the center of $Br_{\gamma}$ line, become negative after that, continue to decrease before increasing to zero again at the last wavelength, $\epsilon_{\rm pol}$ start from zero and increase until the center of $Br_{\gamma}$ line before decreasing again to zero at the last wavelength.

\subsection{MCMC verification}
\label{MCMC}

In order to check the results obtained with $\chi^2$ minimization method and study the coupling of the free parameters between them, we apply the Markov Chain Monte Carlo (MCMC) technique, following the Delayed Rejection and Adaptive Metropolis samplers (DRAM) method \citep{s11222-006-9438-0}. We start around the best four free parameters that we obtained with the $\chi^2$ minimization method for the "param$\_$set3" ($R_{eq}=4\Rsun$, $V_{eq}=300\kms$, $i=85^\circ$ \& $PA_{rot}=250^\circ$). The upper and lower bound was determined as follows: $2\Rsun \leq R_{eq} \leq 5\Rsun$, $250\kms \leq V_{eq} < 450$, $45^\circ \leq i \leq 135^\circ$ \& $200^\circ \leq PA_{rot} \leq 300^\circ$.

Because of the stochasticity of MCMC method, which needs an important number of iterations, we were able to constrain all the free parameters except $V_{\rm eq}$ that we systematically found far above the critical velocity. To fix this problem we deduce the MCMC best parameters by fitting the differential phase and the spectrum data together.

MCMC explores the full posterior distribution using a set of random simulations of SCIROCCO with a frequency of adaptation. The result of the simulations is used for the next step in order to optimize the maximum likelihood. We run MCMC with 600 numbers of simulations and 50 points of adaptation frequency. We do 3 successive runs, starting from the values of the previous run, because we started from non optimized values and the chain needs some time to find the location of the posterior. At the last run, all the simulations are around the maximum likelihood and their average represents well the best solution, while the standard deviation provides the uncertainties. Fig.~\ref{MCMC_Fit_Regulus_Npix=256_Config1} (in appendix) shows, in addition of the covariance matrix, pairs of parameters, with their histogram. The results of this method are close to those of the $\chi^2$ minimization. 
The uncertainties of the parameters estimated by the LM algorithm are $\approx 3\%$ for $R_{\rm eq}$; $\approx 5\%$ for $V_{\rm eq}$; $\approx 2\%$ for $i$ and $\approx 4\%$ for $PA_{rot}$, while the MCMC uncertainties are $\approx 5\%$ for $R_{\rm eq}$ \& $V_{\rm eq}$; $\approx 7\%$ for $i$ and $\approx 8\%$ for $PA_{rot}$. The uncertainties of all the free parameters, except for $PA_{\rm rot}$, are much larger in the MCMC method than in the $\chi^2$ minimisation (between 2 to 5 times), because the MCMC uncertainties are taking into account the coupling between the free parameters. Thus, the MCMC method is supposed to find the real errors, which are considered in this paper (Tab.~\ref{tab_param_Reg1}).

\begin{table*}
\caption{Parameters estimated from a Levenberg-Marquardt fit and their uncertainties from the MCMC fit.}\label{tab_param_Reg1}
\centering
\begin{threeparttable}
\centering
\begin{tabular}{|c|c|c|c|}
  \hline
  \multicolumn{1}{|c|}{\textbf{Parameters}} & \multicolumn{3}{|c|}{\textbf{Regulus}} \\
  \hline
  \textbf{Best-fitting parameter} & \multicolumn{2}{|c|}{\textbf{In the literature}} & \textbf{param$\_$set 3 with MCMC uncertainties}\\
  \hline
  Equatorial radius $R_{\rm eq}$ & \multicolumn{1}{|c}{4.16$\pm$0.08 $\Rsun$ \tnote{1}} & \multicolumn{1}{c|}{$4.21^{+0.07}_{-0.06}$ $\Rsun$ \tnote{2}} & 4.16 $\pm$ 0.24 $\Rsun$ \\
  Equatorial rotation velocity $V_{\rm eq}$ & \multicolumn{1}{|c}{$317^{+3}_{-85}$ $\kms$ \tnote{1}} & \multicolumn{1}{c|}{$337^{+22}_{-33}$ $\kms$ \tnote{2}} & 350$\pm$22 $\kms$ \\
  Rotation-axis inclination angle i & \multicolumn{1}{|c}{$90^{+0}_{-15}$ $^\circ$ \tnote{1}} & \multicolumn{1}{c|}{$86.3^{+1}_{-1.6}$ $^\circ$ \tnote{2}} & 86.4$\pm$6.3$^\circ$ \\
  Rotation-axis position angle $PA_{\rm rot}$ & \multicolumn{1}{|c}{265.5$\pm$2.8$^\circ$ \tnote{1}} & \multicolumn{1}{c|}{$258^{+2}_{-1}$ $^\circ$ \tnote{2}} & 251$\pm$2$^\circ$ \\
  \hline
  N$^\circ$ of free parameters & \multicolumn{1}{|c}{5 \tnote{1}} & \multicolumn{1}{c|}{6 \tnote{2}} & 4 \\
  $\chi^2$ & \multicolumn{1}{|c}{3.35 \tnote{1}} & \multicolumn{1}{|c|}{1.32 \tnote{2}} & \multicolumn{1}{c|}{2.71} \\
  \hline
	\multicolumn{1}{|c|}{\textbf{Fixed parameter}} & \multicolumn{3}{|c|}{\textbf{Value}} \\
  \hline
	\multicolumn{1}{|c|}{Distance d} & \multicolumn{1}{|c}{23.5$\pm$0.4 pc \tnote{1}} & \multicolumn{1}{c|}{24.31$\pm$0.24 pc \tnote{2}} & \multicolumn{1}{c|}{23.759 $\pm$ 0.446 pc \tnote{3}}  \\
	\multicolumn{1}{|c|}{Mass \textit{M}} & \multicolumn{1}{|c}{3.39$\pm$0.24 $\Msun$ \tnote{1}} & \multicolumn{1}{c|}{3.66$^{+0.79}_{-0.28}$ $\Msun$ \tnote{2}} & \multicolumn{1}{c|}{3.8 $\pm$ 0.57 $\Msun$ \tnote{4}} \\
	\multicolumn{1}{|c|}{Surface mean effective temperature $\Tmean$} & \multicolumn{1}{|c|}{12250 K} & \multicolumn{1}{|c|}{12080 K} & \multicolumn{1}{c}{12500 K \tnote{4}} \\
        \multicolumn{1}{|c|}{Gravity-darkening coefficient $\beta$} & \multicolumn{2}{|c|}{Free} & \multicolumn{1}{c}{Estimated} \\
        \multicolumn{1}{|c|}{\textbf{Limb-darkening parameters}} & \multicolumn{2}{|c|}{} & \multicolumn{1}{c|}{}\\
        \multicolumn{1}{|c|}{Turbulent velocity $VT$} & \multicolumn{2}{|c|}{-} & \multicolumn{1}{c|}{2$\kms$}\\
	\multicolumn{1}{|c|}{$\log g$} & \multicolumn{2}{|c|}{$log g_{eff}$=3.5 $cm/s^2$ \tnote{1}} & \multicolumn{1}{c}{4 $cm/s^2$} \\
	\multicolumn{1}{|c|}{Claret $T_{eff}$} & \multicolumn{2}{|c|}{-} & \multicolumn{1}{c}{10500 to 15000 K} \\
	\multicolumn{1}{|c|}{Metallicity $[Fe/H]^e$} & \multicolumn{1}{|c|}{-} & \multicolumn{1}{|c|}{0.0 \tnote{2}} & \multicolumn{1}{c|}{0.0}\\
	\multicolumn{1}{|c|}{Spectral filter} & \multicolumn{2}{|c|}{-} & \multicolumn{1}{c|}{K}\\
	\multicolumn{1}{|c|}{\textbf{Line profile}} & \multicolumn{2}{|c|}{} & \multicolumn{1}{c|}{}\\
	\multicolumn{1}{|c|}{Kind} & \multicolumn{2}{|c|}{-} & \multicolumn{1}{c|}{Kurucz/Synspec}\\
	\multicolumn{1}{|c|}{} & \multicolumn{2}{|c|}{} & \multicolumn{1}{c}{$[10500K,4cm/s^2]$} \\
	\multicolumn{1}{|c|}{Physical characteristics $[T_{eff},log \ g]$} & \multicolumn{2}{|c|}{-} & \multicolumn{1}{c}{to} \\
	\multicolumn{1}{|c|}{} & \multicolumn{2}{|c|}{} & \multicolumn{1}{c}{$[15000K,4cm/s^2]$} \\
	\hline
  \multicolumn{1}{|c|}{\textbf{Derived parameter}} & \multicolumn{3}{|c|}{\textbf{Value}} \\
  \hline
  \multicolumn{1}{|c|}{Equatorial angular diameter $\diameq$} & \multicolumn{1}{|c|}{1.65$\pm$0.02 mas} & \multicolumn{1}{|c|}{1.61$^{+0.03}_{-0.02}$ mas}  & \multicolumn{1}{c}{1.63$\pm$0.09 mas} \\
  \multicolumn{1}{|c|}{Equatorial-to-polar radii $R_{\rm eq}/R_{\rm pol}$} & \multicolumn{1}{|c|}{1.32$\pm$0.04} & \multicolumn{1}{|c|}{1.31$^{+0.05}_{-0.04}$} & \multicolumn{1}{c}{1.35$\pm$0.08} \\
  \multicolumn{1}{|c|}{Critical radius $R_{\rm crit}$} & \multicolumn{1}{|c|}{4.72$\pm$0.04 $\Rsun$} & \multicolumn{1}{|c|}{4.83$^{+0.05}_{-0.04}$ $\Rsun$} & \multicolumn{1}{c}{4.61$\pm$0.53 $\Rsun$} \\
  \multicolumn{1}{|c|}{Critical equatorial rotation velocity $V_{\rm crit}$} & \multicolumn{1}{|c|}{369$^{+14}_{-67}$ $\kms$} &  \multicolumn{1}{|c|}{380$^{+85}_{-32}$ $\kms$} & \multicolumn{1}{c}{396$\pm$46 $\kms$} \\
  \multicolumn{1}{|c|}{$\vsini$} & \multicolumn{1}{|c|}{317$\pm$3 $\kms$} & \multicolumn{1}{|c|}{336$^{+16}_{-24}$ $\kms$} & \multicolumn{1}{c}{349$\pm$60 $\kms$} \\
  \multicolumn{1}{|c|}{$V_{\rm eq}/V_{\rm crit}$} & \multicolumn{1}{|c|}{0.86$\pm$0.03} & \multicolumn{1}{|c|}{0.89$^{0.25}_{0.16}$} & \multicolumn{1}{c}{0.88$\pm$0.05} \\
  \multicolumn{1}{|c|}{$\beta$} & \multicolumn{1}{|c|}{0.25$\pm$0.11} & \multicolumn{1}{|c|}{0.188$^{+0.012}_{-0.029}$} & \multicolumn{1}{c}{0.165$\pm$0.009 \tnote{5}} \\
  \multicolumn{1}{|c|}{Polar effective temperature $T_{\rm pol}$} & \multicolumn{1}{|c|}{$15400\pm1400 $ K} & \multicolumn{1}{|c|}{$14520^{+550}_{-690}$ K} & \multicolumn{1}{c}{$14419 \pm 832$ K} \\ 
  \multicolumn{1}{|c|}{Equatorial effective temperature $T_{\rm eq}$} & \multicolumn{1}{|c|}{$10314\pm1000$ K} & \multicolumn{1}{|c|}{$11010^{+420}_{-520}$ K} & \multicolumn{1}{c}{$10581 \pm 612$ K} \\
  \multicolumn{1}{|c|}{Luminosity $\log L/L_\odot$} & \multicolumn{1}{|c|}{2.540 $\pm$ 0.043} & \multicolumn{1}{|c|}{2.533$^{+0.033}_{-0.037}$} & \multicolumn{1}{c}{2.461 $\pm$ 0.070} \\
  \hline
\end{tabular}  
\begin{tablenotes}
		\footnotesize
		\item[1] \cite{2005ApJ...628..439M};
   		\item[2] \cite{2011ApJ...732...68C};
   		\item[3] \cite{1997A&A...323L..49P};
   		\item[4] \cite{1990A&AS...85.1015M};
   		\item[5] theoretical estimate of $\beta$ from \cite{2011A&A...533A..43E}.
\end{tablenotes}
\end{threeparttable}
\end{table*}

\subsection{Fitting results}
\label{fit_res}

Our results, which come from the photocentre displacements, have relatively greater uncertainties than the results which come from the classical long-baseline and large band interferometry, except for $V_{\rm eq}$, because of the high spectral resolution mode used by AMBER, and for $PA_{\rm rot}$ (with the MCMC method). We confirmed the same value of $PA_{\rm rot}$ with three different methods; directly, with $\chi^2$ minimization and with the MCMC method with good accuracy, and we estimate that our $PA_{\rm rot}$ is fairly reliable.

All our results confirm what it was found before in the literature \citep{2005ApJ...628..439M, 2011ApJ...732...68C}, except for the equatorial velocity $V_{\rm eq}$ which is $\approx10\%$ higher. Despite of this important difference, we argue that our result of $V_{\rm eq}$ is the fairest, because it was deduced from differential phases (photo-center displacement), which are directly related to $V_{\rm eq}$. In addition, our result has been validated by two different minimization methods.

The parameters which are sensitive and which are farther away from a hypothetical case without Von Zeipel effect, are $V_{eq}$ \& $i$. The first parameter that appears with the Von Zeipel effect is the inclination angle "$i$", which gives the true rotation $V_{eq}$ \citep{1972A&A....21..279M}.
The second parameter is the rotation-axis position angle of our star ($PA_{rot}$) that we can deduce, as seen previously, directly from the 2D observed photocentres (see Sec.~\ref{geom_params}). The value adopted for $PA_{\rm rot}$ is $251^\circ$ (West-East rotation direction), is in agreement with the values given by the $\chi^2$/MCMC-fitting ($PA_{rot}=251.11\pm1.82^\circ$).

The gravity darkening coefficient $\beta$ is very important in physics of rotating stars. $\beta=0.25$ is a standard value for stars with radiative envelopes in hydrostatic equilibrium \citep{1924MNRAS..84..665V, 1924MNRAS..84..684V}. We consider it as a first approximation of the surface distribution of the radiative flux with the hypothesis of conservative laws of rotation (centrifugal force obtained from a potential). 
For stars with convective layers, \citet{1967ZA.....65...89L} showed that $\beta\approx0.08$, and \cite{2011ApJ...732...68C} recommend to adopt $\beta\approx0.19$ for the modeling of radiative stars in rotation. 
The insufficient spatial resolution of our observations prevented us to determine the gravity darkening coefficient $\beta$ of Regulus directly, by setting it as a free parameter in $\chi 2$ minimization. Our attempts to constrain $\beta$ using this method produced physically meaningless results, which can be explained taking into account the degeneracy of the solution (see Sec.~\ref{discus}).
In the other hand, we can theoretically estimate it according to the method of \citet{2011A&A...533A..43E}, who propose to use a $\beta$-value adapted to each rotator according to its velocity.
They adopted for Regulus a value of $\beta$ between 0.158 and 0.198 from \cite{2011ApJ...732...68C} results, while using \citet{2011A&A...533A..43E} method, with our $R_{\rm pol}$ \& $R_{\rm eq}$ results (from the Tab.~\ref{tab_param_Reg1}), we deduce that $\beta$ is $\sim 0.17\pm0.1$.

\section{Discussion}
\label{discus}

We have used our SCIRROCCO modeling tool to estimate the effect on $\vec{\epsilon}(\lambda)$ of the rotation axis position angle and inclination, the equatorial radius, the equatorial velocity and the gravitational darkening parameter $\beta$. We have shown that the position angle of the rotation vector can be measured, in direction and orientation, independently from the other parameters as it is the oriented axis of symmetry of the 2D track of $\vec{\epsilon}(\lambda)$  for all possible values of the other parameters. We have computed the limits of accuracy of $\beta$ that we could achieve with our quality of data and this numbers are summarized in table \ref{tab_limit}.

Our estimates show that our SNR is insufficient to give a significant direct constraint on $\beta$ from a fit of our data.

\begin{table}
\centering
\caption{Limiting and achieved accuracy for the parameter $\beta$ of Regulus with our data. The "best possible accuracy" is given for the estimation of a parameter when we assume that all other parameters are known.} \label{tab_limit}
\begin{tabular}{ccc}
\hline \hline
Parameter & Best possible & Accuracy from \\ 
 & accuracy from & MCMC fit of \\
 & $\vec{\epsilon}(\lambda)$ only & $\vec{\epsilon}(\lambda)$ and $s(\lambda)$ \\

\hline \hline

$\beta$ with $\sigma_{\rm R_{\rm eq}}$ & 0.08 & n.a. \\
$\beta$ with $\sigma_{\rm R_{\rm eq}}/R_{\rm eq}=5.7\%$ & 0.3 & n.a. \\

\hline \hline

\end{tabular}
\end{table}

Because of the particular "edge-on" position of Regulus and the symmetry of the equator-on orientation, $(\beta,i)$ should be degenerate \citep{2011ApJ...732...68C}. So from Regulus 2D $\chi^2$ map with "param$\_$set3" and for $0.05 \leq \beta \leq 0.25$ \& $60^\circ \leq i \leq 90^\circ$, we deduced the probability space that shows the degeneracy between stellar parameters $i$ versus $\beta$. 

\begin{figure}
\centering
\includegraphics[width=1.\hsize,draft=false]{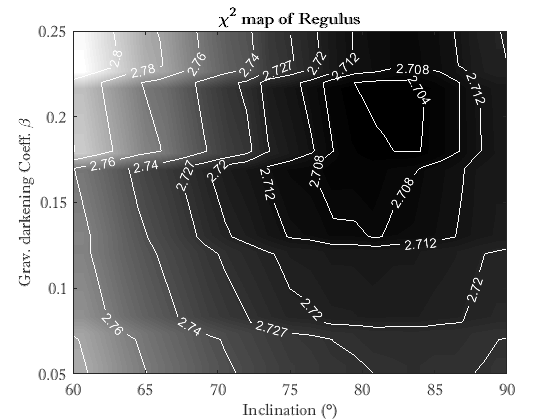}
\caption{$\chi^2$ map of Regulus that shows the degeneracy between the stellar parameters $i$ and $\beta$. The numerical solid white contours represent the $\chi^2$ value. This combination favours values of $0.14 \leq \beta \leq 0.22$ and $77^\circ \leq i \leq 87^\circ$ with a $\chi^2$ value of $2.71$.}
\label{chi2_map}
\end{figure}

From Figure.~\ref{chi2_map}, where we show the Regulus $\chi^2$ map $(\beta,i)$, we can observe an important degeneracy of solution for $\beta$ \& $i$ using our model SCIROCCO. This figure shows a enlarged contour of $\chi^2$, implying an important correlation between $\beta$ \& $i$. 
The value of $\chi^2$ is quasi the same (between 2.704 \& 2.727) in large zone, where $70^\circ\leq i \leq 90^\circ$ and $0.05\leq \beta \leq 0.25$, which makes the accurate determination of $\beta$ very difficult.
The third numerical white solid contour shows the $\chi^2$ zone with a value of $2.712$. Our best model-fitting result from "param$\_$set3" ($i=86.4\pm6.3^\circ$ \& $\beta=0.17\pm0.01$) is in this area too.
The degeneracy ($\beta,i$) in our study is more important than this of \cite{2011ApJ...732...68C} because of the angular resolution quality of the both observations.

It might be possible to improve the constraints by combining our data with the results, or better, the individual measures from previous techniques. For example we could use the angular diameter from previous interferometric measures and better spectra of Regulus (the SNR of the AMBER spectrum is quite poor) for better constraints on the inclination and even of $\beta$. With an accuracy on the radius $\sigma_{\rm R_{\rm eq}}/R_{\rm eq}=1.5\%$ as in \cite{2005ApJ...628..439M,2011ApJ...732...68C} measures, our data would yield $\sigma_\beta\approx 0.15$ and a gain in SNR$\sim$3 would be enough to achieve $\sigma_\beta\approx 0.05$. For high spectral resolution differential phase measurements, the accuracy on the photocentre increases proportionally with angular diameter, aperture size and baseline length \citep{1989dli..conf..249P}. With larger apertures (the Unit Telescopes) or larger baselines and shorter wavelengths (CHARA in the visible with a high spectral resolution instrument) and even with smaller and fainter sources, such a gain in SNR seems quite accessible. A source with an inclination closer to 45$^\circ$ would also provide a larger signal in the polar direction (Fig.~\ref{horseshoes}d). This opens excellent perspectives for future observations of rapid rotators with high spectral resolution differential interferometry.

From the Be's catalog of \cite{2005A&A...440..305F}, we found $\sim 50\%$ of stars with $m_V=$2.33 - 7.3, which are corresponding to the selection criteria of $30 \leq i \leq 60$ and $v/v_c \leq 70\%$, and that could be studied by the method that we propose in this paper.

\section{Conclusions }
\label{conclusions}

We have presented the differential interferometry data obtained on the rapid rotator Regulus with the high spectral resolution mode of the VLTI instrument AMBER. We have seen that, for K band observations with the VLTI baselines, this target much smaller than the angular resolution $(\lambda/B)$ is not enough resolved for the Rayleigh criteria and for imaging, as all closure phases are equal to zero. It is also not enough resolved for oblateness estimations from absolute visibility measurements, because the visibility is equal to 1 within the error bars for the baselines close to the polar direction and we can only give an upper limit for the polar diameter that is smaller than the equatorial diameter. 
We have therefore concentrated on the interpretation and model fitting of the differential phases that, on this source much smaller than the standard resolution limit $\lambda/B$, can all be reduced to the vectorial displacement of the photocentre $\vec{\epsilon}(\lambda)$ in the spectral channel $\lambda$ with respect to the photocentre of the target in the continuum.
Our data, corresponding to $\sim$30 mn ($\sim$25 mn for the night 2014/03/10 \& $\sim$40 mn for the night 2014/03/12) of open shutter observations on Regulus, yields a typical error per spectral measure (half spectral channel) of 30 $\mu as$ both for the $\alpha$ and $\delta$ components of the vector $\vec{\epsilon}(\lambda)$ and for its polar and equatorial components obtained after a direct computation of the rotation axis position angle.

We have basically confirmed the previous interferometric and spectroscopic determinations of the fundamental parameters of Regulus, with our quite different data set and different constraints on the physical parameters. Our $350\pm22 \kms$ velocity measurement is compatible within errors with this used by \cite{2011ApJ...732...68C}, but significantly higher than earlier estimates. This leads to a rotation 88\% of the critical velocity. Our $PA_{\rm rot}$ measurement of $251\pm2^\circ$, which is only deduced from the vectorial photocentre ($\phidiff$), is little bit different of this deduced from squared visibilities, closure phase and triple amplitudes of CHARA/MIRC by \cite{2011ApJ...732...68C}. The vectorial photocentre is very sensitive to $PA_{\rm rot}$ parameter.

We don't claim at all that our values of $\beta$ are conclusive, because we are not able to constrain the gravity darkening coefficient from our current data, which are relatively noisy with marginal angular resolution, and without forgetting the fact that Regulus is a edge-on star.

Despite the fact that the star was marginally resolved with our observations, we were able, for the first time, to constrain independently (from the $\epsilon_{\alpha}$ \& $\epsilon_{\delta}$) several fundamental stellar parameters, as the $PA_{\rm rot}$ with low uncertainties.
This method can be applied to stars which can be only marginally resolved or not angularly resolved at all, because of available baseline lenghts, and especially for rotators with inclinaison angles around of $45^\circ$ and a good SNR.


\bibliographystyle{mnras}
\bibliography{Biblio_Reg}


\appendix

\section{MCMC plots and study of correlations}
\label{MCMC_figs}

\begin{figure*}
\centering
\includegraphics[width=1.\hsize,draft=false]{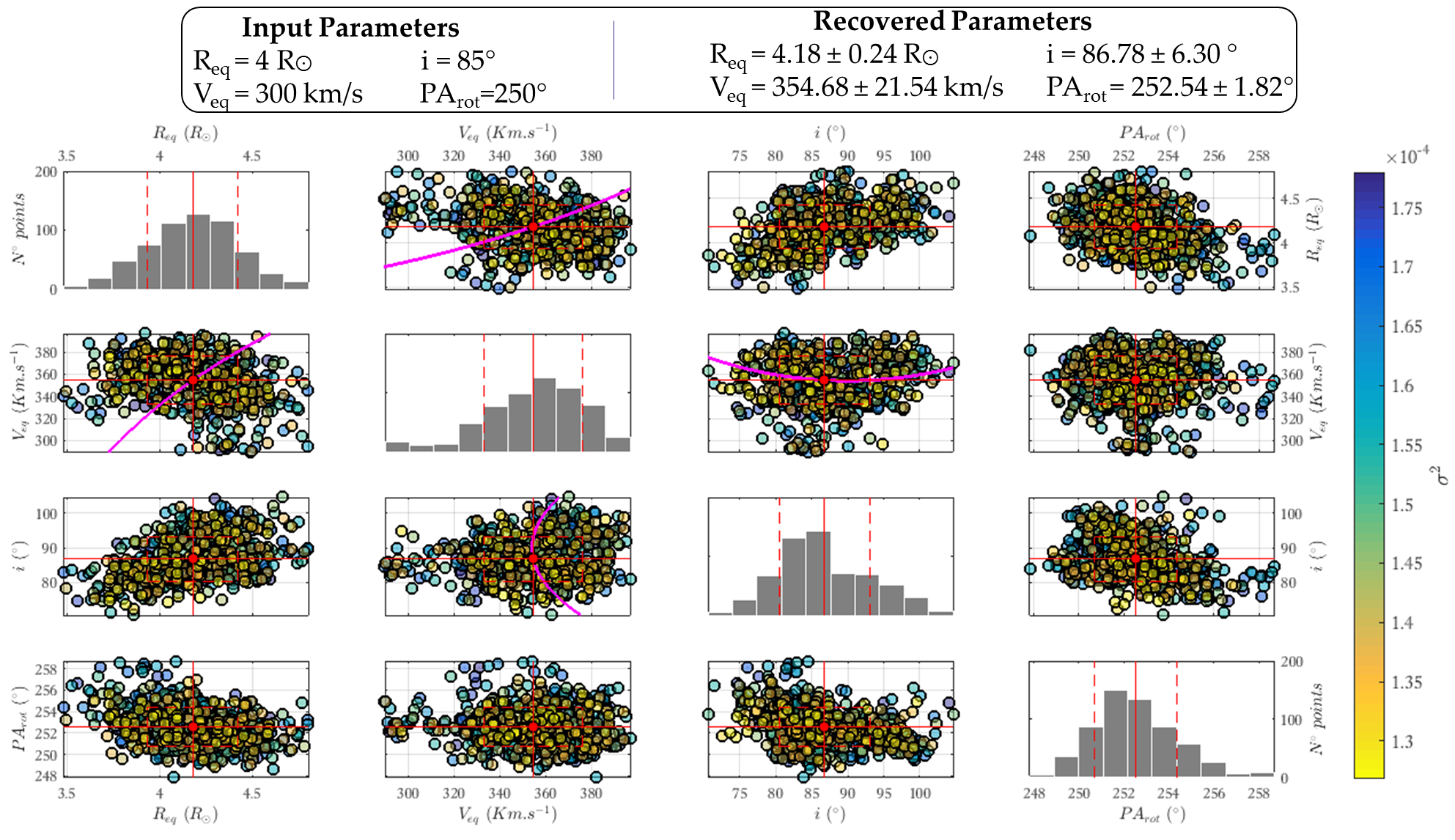}
\caption{MCMC (DRAM) covariance matrix distribution results for the four free parameters ($R_{eq}$, $V_{eq}$, $i$ and $PA_{rot}$) of Regulus. The red point and line show the best recovered parameters, the average of the last MCMC run. The scatter plots show the projected two-dimensional distributions of the projected covariance matrix (coloured points) two by two parameters. The colour bar represents the distribution of the points around the average following the variance $\sigma^2$. The histograms show the projected one-dimensional distributions with solid red lines representing the best recovered parameters and dashed red line the uncertainties. From top to bottom and left to right, the panels show the equatorial radius $R_{\rm eq}$, equatorial rotation velocity $V_{\rm eq}$, rotation-axis inclination angle $i$ and rotation-axis position angle $PA_{\rm rot}$; similarly, by symmetry, to the histogram plots. The hypothetical behaviour of $V_{\rm eq}=f(i)$ is shown by a magenta continuous line following $\vsini$; also $R_{\rm eq}=f(R_{\rm eq})$, which behaves theoretically as $R_{\rm eq}=R_{\rm pol}\left(1-V^2_{\rm eq}R_{\rm p}/2GM\right)$, is a magenta line. We observe that the distribution of the points is around those behaviours in both cases.} 
\label{MCMC_Fit_Regulus_Npix=256_Config1}
\end{figure*}

Fig.~\ref{MCMC_Fit_Regulus_Npix=256_Config1} shows the covariance matrix: pairs of parameters, with their histogram, that were obtained by the MCMC method.

At first sight, among all four parameters, only the equatorial velocity $V_{eq}$ and the inclination $i$ are strongly correlated, because they are both correlated with $\vsini$. However, quantitatively and using the correlation coefficient $\rho$ which is defined for two scalar quantities $A$ and $B$ as $\rho(A,B)=cov(A,B)/(\sigma_A*\sigma_B)$, where $cov$ is the covariance and $\sigma$ the standard deviation, we found, in descending order, $\rho(R_{\rm eq},i)=0.4247$, $\rho(R_{\rm eq},PA_{\rm rot})=-0.3379$, $\rho(i,PA_{\rm rot})=-0.3228$, $\rho(R_{\rm eq},V_{\rm eq})=-0.3044$, $\rho(V_{\rm eq},i)=0.1124$ and $\rho(V_{\rm eq},PA_{\rm rot})=-0.0730$.
When $\rho=1$ it means full correlation, while $\rho=0$ means no correlation. When $\rho>0$, the correlation is proportional ($A$ and $B$ increase or decrease together), and when $\rho<0$ the correlation is inversely proportional ($A$ increases when $B$ decreases and vice versa), e.g. the case of $R_{\rm eq}$ and $V_{\rm eq}$, because of angular momentum conservation. Note also that the correlation coefficients of all our free MCMC parameters are symmetric (i.e. $\rho(A,B)=\rho(B,A)$).

\section{SCIROCCO's deduced parameters}
\label{scirocco_ded}

Tab.~\ref{tab_eq_Reg1} summarizes all the equations of the deduced parameters used by SCIROCCO, where $G$ is the gravitational constant, $\sigma$ the Stefan-Boltzmann constant and $\theta$ the co-latitude (note that for the critical equatorial rotation velocity, the Eddington factor can be ignored for Regulus, because it presents low luminosity).

\begin{table}
\begin{minipage}{87mm}
\centering
\caption{SCIROCCO's deduced parameter formulae.} \label{tab_eq_Reg1}
\begin{threeparttable}
\begin{tabular}{l|c}
\hline \hline
Parameter & Formula\\
\hline \hline
Angular diameter & $\diameq=2\frac{R_{\rm eq}}{d}\frac{180}{\pi}36\times10^5$ (3)\\
\hline
Equ-to-pol. radii & $\frac{R_{\rm pol}}{R_{\rm eq}}=\left(1+\frac{V^2_{\rm eq}R_{\rm eq}}{2GM}\right)^{-1}$ (1,2,3)\\
\hline
Critical radius & $R_{\rm crit}=3\frac{R_{\rm pol}}{2}$ (1,2,3)\\
\hline
Crit. equa. rot. velocity & $V_{\rm crit}=\sqrt{\frac{GM}{R_{\rm crit}}}$ (1,2,3)\\
\hline
Latitudinal $T_\mathrm{eff}$ & $T_\mathrm{eff}(\theta)=\left (\frac{C}{\sigma} \right )^{0.25} g_\mathrm{eff}^{\beta}(\theta)$ (1,2,3)\\
 & where,\\ 
 & the constant $C=\frac{\sigma \Tmean^4 S_\star}{\int g_\mathrm{eff}^{4\beta}(\theta)\,dS}$,\\
 & and $T_\mathrm{eq}=T_\mathrm{eff}(90^\circ)$\\
 & \& $T_\mathrm{pol}=T_\mathrm{eff}(0^\circ(+180^\circ))$\\
\hline
Luminosity & $\log L/L_\odot=\log\left(\frac{S_\star\sigma\Tmean^4}{4\pi R_\odot^2\sigma T_\odot^4}\right)$ (3)\\
\hline \hline
\end{tabular}
\begin{tablenotes}
		\footnotesize
		$(1)$ {\citet{2002A&A...393..345D}}
   	$(2)$ {\citet{2014A&A...569A..45H}}\\
   	$(3)$ \citet{Massi2015}
\end{tablenotes}
\end{threeparttable}
\end{minipage}
\end{table}

\section{All the fit results}
\label{all_fit_res}

Table \ref{tab_param_Reg_plus} summarizes all the fit results that we found for different parameter sets and methods.

\begin{table*}
\caption{Parameters with their uncertainties estimated from Levenberg-Marquardt and MCMC fits for all the parameter sets.}\label{tab_param_Reg_plus}
\centering
\begin{sideways}
\centering
\begin{threeparttable}
\centering
\begin{tabular}{|c|c|c|c|c|c|c|c|}
  \hline
  \multicolumn{1}{|c|}{\textbf{Parameters}} & \multicolumn{7}{|c|}{\textbf{Regulus}} \\
  \hline
  \multicolumn{1}{|c|}{} & \multicolumn{7}{|c|}{\textbf{Comparison of best-fitting values between the four parameter sets}} \\
  \cline{2-8}
  \textbf{Best-fitting parameter} & \multicolumn{2}{|c|}{\textbf{In the literature}} & \textbf{param$\_$set 1} & \textbf{param$\_$set 2} & \textbf{param$\_$set 3} & \textbf{param$\_$set 4} & \textbf{MCMC}\\
  \hline
  Equatorial radius $R_{\rm eq}$ & \multicolumn{1}{|c}{4.16$\pm$0.08 $\Rsun$ \tnote{1}} & \multicolumn{1}{c|}{$4.21^{+0.07}_{-0.06}$ $\Rsun$ \tnote{2}} & 4.16 $\pm$ 0.12 $\Rsun$ & 4.00 $\pm$ 0.18 $\Rsun$ & 4.16 $\pm$ 0.11 $\Rsun$ & 4.00 $\pm$ 0.17 $\Rsun$ & 4.18 $\pm$ 0.24 $\Rsun$\\
  Equatorial rotation velocity $V_{\rm eq}$ & \multicolumn{1}{|c}{$317^{+3}_{-85}$ $\kms$ \tnote{1}} & \multicolumn{1}{c|}{$337^{+22}_{-33}$ $\kms$ \tnote{2}} & 350$\pm$15 $\kms$ & 358$\pm$16 $\kms$ & 350$\pm$18 $\kms$ & 358$\pm$13 $\kms$ & 355$\pm$22 $\kms$ \\
  Rotation-axis inclination angle i & \multicolumn{1}{|c}{$90^{+0}_{-15}$ $^\circ$ \tnote{1}} & \multicolumn{1}{c|}{$86.3^{+1}_{-1.6}$ $^\circ$ \tnote{2}} & 86.4$\pm$1.8$^\circ$ & 86.2$\pm$2.3$^\circ$ & 86.4$\pm$2.1$^\circ$ & 86.2$\pm$1.6$^\circ$ & 86.8$\pm$6.3$^\circ$ \\
  Rotation-axis position angle $PA_{\rm rot}$ & \multicolumn{1}{|c}{265.5$\pm$2.8$^\circ$ \tnote{1}} & \multicolumn{1}{c|}{$258^{+2}_{-1}$ $^\circ$ \tnote{2}} & 251$^\circ$ & 251$^\circ$ & 251$\pm$10$^\circ$ & 251$\pm$5$^\circ$ & 253$\pm$2$^\circ$ \\
  \hline
  N$^\circ$ of free parameters & \multicolumn{1}{|c}{5 \tnote{1}} & \multicolumn{1}{c|}{6 \tnote{2}} & 3 & 3 & 4 & 4 & 4 \\
  $\chi^2$ & \multicolumn{1}{|c}{3.35 \tnote{1}} & \multicolumn{1}{c|}{1.32 \tnote{2}} & 2.70 & 2.70 & 2.71 & 2.71 & - \\
  \hline
	\multicolumn{1}{|c|}{\textbf{Fixed parameter}} & \multicolumn{7}{|c|}{\textbf{Value}} \\
  \hline
	\multicolumn{1}{|c|}{Distance d} & \multicolumn{1}{|c}{23.5$\pm$0.4 pc \tnote{1}} & \multicolumn{1}{c|}{24.31$\pm$0.24 pc \tnote{2}} & \multicolumn{5}{c|}{23.759 $\pm$ 0.446 pc \tnote{3}}  \\
	\multicolumn{1}{|c|}{Mass \textit{M}} & \multicolumn{1}{|c}{3.39$\pm$0.24 $\Msun$ \tnote{1}} & \multicolumn{1}{c|}{3.66$^{+0.79}_{-0.28}$ $\Msun$ \tnote{2}} & \multicolumn{5}{c|}{3.8 $\pm$ 0.57 $\Msun$ \tnote{4}} \\
	\multicolumn{1}{|c|}{Surface mean effective temperature $\Tmean$} & \multicolumn{1}{|c|}{12250 K} & \multicolumn{1}{|c|}{12080 K} & \multicolumn{5}{c}{12500 K \tnote{4}} \\
        \multicolumn{1}{|c|}{Gravity-darkening coefficient $\beta$} & \multicolumn{2}{|c|}{Free} & \multicolumn{1}{c}{Estimated} & \multicolumn{1}{c}{0.25} & \multicolumn{1}{c}{Estimated} & \multicolumn{1}{c}{0.25} & \multicolumn{1}{c}{Estimated}\\
        \multicolumn{1}{|c|}{\textbf{Limb-darkening parameters}} & \multicolumn{2}{|c|}{} & \multicolumn{4}{c|}{}\\
        \multicolumn{1}{|c|}{Turbulent velocity $VT$} & \multicolumn{2}{|c|}{-} & \multicolumn{5}{c|}{2$\kms$}\\
	\multicolumn{1}{|c|}{$\log g$} & \multicolumn{2}{|c|}{$log g_{eff}$=3.5 $cm/s^2$ \tnote{1}} & \multicolumn{4}{c}{4 $cm/s^2$} \\
	\multicolumn{1}{|c|}{Claret $T_{eff}$} & \multicolumn{2}{|c|}{-} & \multicolumn{5}{c}{10500 to 15000 K} \\
	\multicolumn{1}{|c|}{Metallicity $[Fe/H]^e$} & \multicolumn{1}{|c|}{-} & \multicolumn{1}{|c|}{0.0 \tnote{2}} & \multicolumn{5}{c|}{0.0}\\
	\multicolumn{1}{|c|}{Spectral filter} & \multicolumn{2}{|c|}{-} & \multicolumn{5}{c|}{K}\\
	\multicolumn{1}{|c|}{\textbf{Line profile}} & \multicolumn{2}{|c|}{} & \multicolumn{5}{c|}{}\\
	\multicolumn{1}{|c|}{Kind} & \multicolumn{2}{|c|}{-} & \multicolumn{1}{c|}{Kurucz/Synspec} & \multicolumn{1}{c|}{Kurucz/Synspec} & \multicolumn{1}{c|}{Kurucz/Synspec} & \multicolumn{1}{c|}{Kurucz/Synspec} & \multicolumn{1}{c|}{Kurucz/Synspec}\\
	\multicolumn{1}{|c|}{} & \multicolumn{2}{|c|}{} & \multicolumn{1}{c}{$[10500K,4cm/s^2]$} & \multicolumn{1}{c}{$[9500K,4cm/s^2]$} & \multicolumn{1}{c}{$[10500K,4cm/s^2]$} & \multicolumn{1}{c}{$[9500K,4cm/s^2]$} & \multicolumn{1}{c}{$[10500K,4cm/s^2]$} \\
	\multicolumn{1}{|c|}{Physical characteristics $[T_{eff},log \ g]$} & \multicolumn{2}{|c|}{-} & \multicolumn{1}{c}{to}  & \multicolumn{1}{c}{to}  & \multicolumn{1}{c}{to}  & \multicolumn{1}{c}{to}  & \multicolumn{1}{c}{to} \\
	\multicolumn{1}{|c|}{} & \multicolumn{2}{|c|}{} & \multicolumn{1}{c}{$[15000K,4cm/s^2]$}  & \multicolumn{1}{c}{$[15000K,4cm/s^2]$}  & \multicolumn{1}{c}{$[15000K,4cm/s^2]$}  & \multicolumn{1}{c}{$[15000K,4cm/s^2]$}  & \multicolumn{1}{c}{$[15000K,4cm/s^2]$} \\
	\hline
  \multicolumn{1}{|c|}{\textbf{Derived parameter}} & \multicolumn{7}{|c|}{\textbf{Value}} \\
  \hline
  \multicolumn{1}{|c|}{Equatorial angular diameter $\diameq$} & \multicolumn{1}{|c|}{1.65$\pm$0.02 mas} & \multicolumn{1}{|c|}{1.61$^{+0.03}_{-0.02}$ mas} & \multicolumn{1}{c}{1.63$\pm$0.05 mas} & \multicolumn{1}{c}{1.57$\pm$0.07 mas} & \multicolumn{1}{c}{1.63$\pm$0.04 mas} & \multicolumn{1}{c}{1.57$\pm$0.07 mas} & \multicolumn{1}{c}{1.64$\pm$0.09 mas} \\
  \multicolumn{1}{|c|}{Equatorial-to-polar radii $R_{\rm eq}/R_{\rm pol}$} & \multicolumn{1}{|c|}{1.32$\pm$0.04} & \multicolumn{1}{|c|}{1.31$^{+0.05}_{-0.04}$} & \multicolumn{1}{c}{1.35$\pm$0.04} & \multicolumn{1}{c}{1.35$\pm$0.06} & \multicolumn{1}{c}{1.35$\pm$0.03} & \multicolumn{1}{c}{1.35$\pm$0.06} & \multicolumn{1}{c}{1.36$\pm$0.08} \\
  \multicolumn{1}{|c|}{Critical radius $R_{\rm crit}$} & \multicolumn{1}{|c|}{4.72$\pm$0.04 $\Rsun$} & \multicolumn{1}{|c|}{4.83$^{+0.05}_{-0.04}$ $\Rsun$} & \multicolumn{1}{c}{4.62$\pm$0.26 $\Rsun$} & \multicolumn{1}{c}{4.43$\pm$0.39 $\Rsun$} & \multicolumn{1}{c}{4.61$\pm$0.24 $\Rsun$} & \multicolumn{1}{c}{4.43$\pm$0.38 $\Rsun$} & \multicolumn{1}{c}{4.60$\pm$0.53 $\Rsun$} \\
  \multicolumn{1}{|c|}{Critical equatorial rotation velocity $V_{\rm crit}$} & \multicolumn{1}{|c|}{369$^{+14}_{-67}$ $\kms$} &  \multicolumn{1}{|c|}{380$^{+85}_{-32}$ $\kms$} &  \multicolumn{1}{c}{396$\pm$22 $\kms$} & \multicolumn{1}{c}{404$\pm$36 $\kms$} & \multicolumn{1}{c}{396$\pm$20 $\kms$} & \multicolumn{1}{c}{404$\pm$35 $\kms$} & \multicolumn{1}{c}{397$\pm$46 $\kms$} \\
  \multicolumn{1}{|c|}{$\vsini$} & \multicolumn{1}{|c|}{317$\pm$3 $\kms$} & \multicolumn{1}{|c|}{336$^{+16}_{-24}$ $\kms$} & \multicolumn{1}{c}{349$\pm$26 $\kms$} & \multicolumn{1}{c}{357$\pm$29 $\kms$}  & \multicolumn{1}{c}{349$\pm$31 $\kms$}  & \multicolumn{1}{c}{357$\pm$23 $\kms$} & \multicolumn{1}{c}{354$\pm$60 $\kms$}\\
  \multicolumn{1}{|c|}{$V_{\rm eq}/V_{\rm crit}$} & \multicolumn{1}{|c|}{0.86$\pm$0.03} & \multicolumn{1}{|c|}{0.89$^{0.25}_{0.16}$} & \multicolumn{1}{c}{0.88$\pm$0.04} & \multicolumn{1}{c}{0.89$\pm$0.03} & \multicolumn{1}{c}{0.88$\pm$0.04} & \multicolumn{1}{c}{0.89$\pm$0.03} & \multicolumn{1}{c}{0.89$\pm$0.05}\\
  \multicolumn{1}{|c|}{$\beta$} & \multicolumn{1}{|c|}{0.25$\pm$0.11} & \multicolumn{1}{|c|}{0.188$^{+0.012}_{-0.029}$} & \multicolumn{1}{c}{0.165$\pm$0.005 \tnote{5}} & \multicolumn{1}{c}{0.163$\pm$0.007 \tnote{5}} & \multicolumn{1}{c}{0.165$\pm$0.004 \tnote{5}} & \multicolumn{1}{c}{0.163$\pm$0.007 \tnote{5}} & \multicolumn{1}{c}{0.162$\pm$0.009 \tnote{5}} \\
  \multicolumn{1}{|c|}{Polar effective temperature $T_{\rm pol}$} & \multicolumn{1}{|c|}{$15400\pm1400 $ K} & \multicolumn{1}{|c|}{$14520^{+550}_{-690}$ K} & \multicolumn{1}{c}{$14420 \pm 815$ K} &  \multicolumn{1}{c}{$15540 \pm 686$ K} & \multicolumn{1}{c}{$14419 \pm 738$ K} & \multicolumn{1}{c}{$15537 \pm 671$ K} & \multicolumn{1}{c}{$14410 \pm 827$ K} \\ 
  \multicolumn{1}{|c|}{Equatorial effective temperature $T_{\rm eq}$}  & \multicolumn{1}{|c|}{$10314\pm1000$ K} & \multicolumn{1}{|c|}{$11010^{+420}_{-520}$ K} & \multicolumn{1}{c}{$10580 \pm 598$ K} & \multicolumn{1}{c}{$9460 \pm 418$ K} & \multicolumn{1}{c}{$10581 \pm 542$ K} & \multicolumn{1}{c}{$9463 \pm 409$ K} & \multicolumn{1}{c}{$10590 \pm 608$ K} \\
  \multicolumn{1}{|c|}{Luminosity $\log L/L_\odot$} & \multicolumn{1}{|c|}{2.540 $\pm$ 0.043} & \multicolumn{1}{|c|}{2.533$^{+0.033}_{-0.037}$} & \multicolumn{1}{c}{2.463 $\pm$ 0.034} & \multicolumn{1}{c}{2.428 $\pm$ 0.055} & \multicolumn{1}{c}{2.461 $\pm$ 0.031} & \multicolumn{1}{c}{2.428 $\pm$ 0.053} & \multicolumn{1}{c}{2.464 $\pm$ 0.069} \\
  \hline
\end{tabular}  
\begin{tablenotes}
		\footnotesize
		\item[1] \cite{2005ApJ...628..439M};
   		\item[2] \cite{2011ApJ...732...68C};
   		\item[3] \cite{1997A&A...323L..49P};
   		\item[4] \cite{1990A&AS...85.1015M};
   		\item[5] theoretical estimate of $\beta$ from \cite{2011A&A...533A..43E}.
\end{tablenotes}
\end{threeparttable}
\end{sideways}
\end{table*}

\section*{ACKNOWLEDGEMENTS }
\label{acknowledgements}
      This research made use of the SIMBAD database, operated at the CDS, Strasbourg, France, and of the NASA Astrophysics Data System Abstract Service. The author, M. Hadjara, acknowledges support from the scientific French association PSTJ \footnote{\url{http://www.pstj.fr/}} for its official host agreement, the Lagrange and OCA for computer server support. This research made use of the Jean-Marie Mariotti Center \texttt{SearchCal} service \footnote{Available at \url{http://www.jmmc.fr/searchcal}} codeveloped by Lagrange and IPAG, and of the CDS Astronomical Databases SIMBAD and VIZIER \footnote{Available at \url{http://cdsweb.u-strasbg.fr/}}. This research made use of the \texttt{AMBER data reduction package} of the Jean-Marie Mariotti Center\footnote{Available at \url{http://www.jmmc.fr/amberdrs}}. Special thanks go to the project's grant ALMA-CONICYT N$^\circ$ 31150002 and the PI, Keiichi Ohnaka, who supported this work.


\end{document}